\documentclass[10pt,twocolumn]{article}

\usepackage[margin=0.75in,columnsep=0.25in]{geometry}
\usepackage[T1]{fontenc}
\usepackage[utf8]{inputenc}
\usepackage{microtype}

\usepackage{amsmath}
\usepackage{amssymb}
\usepackage{dsfont}
\usepackage{graphicx}
\usepackage{tikz}

\usepackage{adjustbox}

\usepackage{booktabs}

\usepackage{pifont} 
\newcommand{\cmark}{\ding{51}} 
\newcommand{\xmark}{\ding{55}} 
\usepackage{multirow}
\usepackage{array}
\usepackage{makecell}
\usepackage{tabularx}
\usepackage{adjustbox}
\usepackage{etoolbox}

\usepackage{amsmath}
\usepackage{amssymb}
\usepackage{booktabs} 
\usepackage{multirow}
\usepackage{subcaption}
\usepackage{svg}
\usepackage{array}
\usepackage{makecell}
\usepackage{paralist}

\usepackage{subcaption}

\usepackage[hidelinks]{hyperref}

\renewcommand{\arraystretch}{1.05}

\AtBeginEnvironment{table}{\small\centering}
\AtBeginEnvironment{table*}{\small\centering}

% ...
% ...

\title{When the World Lies: \\  Backdoor Attacks on Latent World Models for Downstream Control}

\author{%
\small
\textbf{Roberto Riaño}$^{1,2}$ \quad
\textbf{Gorka Abad}$^{3}$ \quad
\textbf{Stjepan Picek}$^{1,4}$ \quad
\textbf{Aitor Urbieta}$^{2}$
\\[0.5em]
\footnotesize
$^{1}$Radboud University, The Netherlands
\quad
$^{2}$IKERLAN Technology Research Centre, Spain
\\\footnotesize
$^{3}$University of Bergen, Norway
\quad
$^{4}$University of Zagreb, Croatia
\\[0.4em]
\scriptsize
\texttt{roberto.rianohidalgo@ru.nl} \quad
\texttt{gorka.abad@uib.no}\quad
\texttt{stjepan.picek@ru.nl} \quad
\texttt{aurbieta@ikerlan.es}
}

\date{}

\begin{document}
\maketitle

\begin{abstract}

Pretrained world models, learned simulators that encode an observation into a latent state and predict how it evolves under actions, are beginning to be reused as off-the-shelf dynamics backbones for control, like pretrained encoders and language models are reused today. We show that this reuse opens a supply-chain backdoor: an adversary who controls only a released checkpoint can hijack the downstream controller, even though the victim trains and evaluates entirely on clean data and never sees the trigger. The attack encodes no explicit trigger-to-action rule. Instead, the poisoned model routes trigger-bearing observations into a chosen latent region and reshapes the local dynamics there, so that the victim's own optimization (Dreamer-style actor training in imagination, or MPC/CEM planning over predicted futures) re-discovers the attacker's target action on its own. Across several control tasks and trigger families, the trigger steers the controller's action toward the attacker's target, controlling every action dimension and hijacking 100\% of triggered steps on the strongest settings. The checkpoint still passes the clean-data diagnostics a victim would run before deployment, with clean-task success retaining at least $\sim$75\%. The effect is temporally gated: it appears only while the trigger is present and disappears when the trigger is removed. Trigger-blind repair is budget-dependent: moderate clean fine-tuning can preserve clean utility while leaving the triggered failure intact, whereas sufficiently aggressive adaptation can remove it only after substantially degrading clean control. The world-model backbone itself is therefore an emerging and underexamined attack surface for control. The full code and artifacts are available in our repository.
\end{abstract}


%

\section{Introduction}
\label{sec:intro}

In the same way that Large Language Models (LLMs) now serve as reusable backbones for many downstream tasks, world models are emerging as reusable backbones for decision-making systems~\cite{ha2018worldmodels,LeCun2022APT,bommasani2021foundation}. A world model is a learned simulator: it encodes an observation, such as an image, into an internal state and predicts how that state will evolve after an action. A downstream controller can then use these predictions to decide what to do, either by training on imagined futures, as in Dreamer-style agents~\cite{hafner2025dreamer}, or by searching over possible action sequences, as in model-predictive control (MPC)~\cite{le2024lewm}.

Today, many world-model systems are still small enough to train in a research lab. However, this is unlikely to remain true as the field moves toward larger video-based and control-oriented world models, such as Genie~\cite{bruce2024genie}, GAIA-1~\cite{hu2023gaia1}, and V-JEPA~\cite{bardes2024vjepa}. As these models become more expensive to train, downstream users will increasingly rely on externally supplied checkpoints rather than training their own models from scratch. This creates a natural supply-chain setting: a user may trust a pretrained world model as the dynamics backbone for a controller, much as pretrained LLM checkpoints are reused today~\cite{bommasani2021foundation}. This reuse pattern raises a security question that has received little attention: what if the supplied world model is malicious? Unlike a classifier or static feature extractor, a world model does not directly emit the downstream decision. It produces action-conditioned latent dynamics that a separate controller later optimizes. A trigger-to-anchor mapping can therefore control where this optimization starts, but not which action an unknown downstream controller selects. The malicious checkpoint must also shape the local dynamics so that the victim's own optimization reconstructs the attacker's target action. This is the central difference from conventional latent backdoors: the backdoor has to survive an intervening optimization loop rather than directly select an output.

This setting also differs from many reinforcement-learning backdoors, where the attacker tampers with rewards, observations, or demonstrations during the victim's training process~\cite{kiourti2020trojdrl,yang2019design,wang2021backdoorl}. We study a checkpoint-only supply-chain attack. The attacker controls the world model before release, while the victim controls the downstream controller, task objective, clean data, and evaluation procedure. The trigger is never shown during victim training. The compromise must therefore be embedded in the world model's dynamics and later propagated through a controller that the attacker does not control.

We study this supply-chain threat in two common uses of world models: policy training from imagined futures and planning by search over predicted futures. In both cases, the attacker poisons the world model before release. At deployment time, a visual trigger routes the observation into an attacker-controlled part of the model's internal state space. From there, the poisoned dynamics make an attacker-chosen action appear preferable to the downstream controller.

Our results show that this attack can hijack downstream control while preserving clean behavior. We evaluate the threat across both actor-training and planner-based settings, multiple control environments, and several trigger families. The attack remains difficult to identify using victim-side diagnostics such as clean prediction error, clean task performance, latent similarity, and prediction drift. We also find that common defenses weaken the directed hijack but do not fully remove the triggered effect. We make the following contributions:
\begin{compactenum}
\item \textbf{A world-model supply-chain threat model.} We study an attacker who controls only the released dynamics checkpoint, while the victim controls the downstream controller, clean data, task objective, and evaluation procedure. This captures a realistic reuse scenario for pretrained world models.

\item \textbf{Two backdoor attacks for downstream control.} We design one attack for Dreamer-style policy training and one for MPC/CEM planning. Both attacks poison the world model itself, so the compromised behavior transfers to the controller trained or executed on top.

\item \textbf{A trigger-gated action hijack.} The backdoor steers the controller's action toward an attacker-chosen target only when the trigger is present; when it is removed, clean behavior is largely preserved, making the attack temporally localized rather than a general degradation.

\item \textbf{Evaluation across controllers, tasks, and triggers.} We evaluate the attacks across multiple control settings and show that different trigger families can activate the same underlying vulnerability.

\item \textbf{Stealth and durability analysis.} We study whether the attack can be detected or removed using diagnostics and defenses available to the victim. Clean-data checks do not reliably separate poisoned from clean checkpoints; evaluated repair-time defenses weaken the directed action but cannot remove the backdoor, which persists as a durable, trigger-gated denial-of-service. A deployment-time detector can flag the trigger, but cannot, in a control loop, restore correct behavior.
\end{compactenum}
 
\section{Background}
\label{sec:background}

A \emph{latent} world model adds one more representation step. The encoder $e_\theta$ maps an observation $o_t$ to a compact latent state $z_t = e_\theta(o_t)$, and the latent transition model $f_\theta$ predicts the next latent under an action, $z_{t+1} = f_\theta(z_t, a_t)$. The downstream controller, whether a learned policy or a planner, never sees raw observations during training or planning; its view of the environment is the trajectory of latents the world model predicts. These models are primarily used in one of two ways: actor training or latent space planning.

\noindent{\textbf{Dreamer-style actor training}}.
DreamerV3~\cite{hafner2025dreamer} trains an actor and a value head from \emph{imagined} states produced by the world model, without running the actor in the real environment during training. At each training step, the actor proposes an action $a_t$ given the latent state $z_t$. The world model rolls the latent forward $H$ steps under those actions, and a reward head predicts the reward at each imagined latent state. Then the actor is updated to maximize the imagined return. The actor's training signal is therefore generated entirely by the world model. 

\noindent{\textbf{MPC/CEM planning in latent space}}. LeWorldModel~\cite{le2024lewm} represents the second paradigm: no actor is trained at all. Instead, at each control step, a Model-Predictive Control loop with Cross-Entropy Method sampling (MPC/CEM) interacts directly with the world model. The planner maintains a Gaussian over action sequences of length $H$ and draws a batch of candidate sequences. Each of these sequences is rolled through the latent dynamics to compute a goal-conditioned cost, keeping the lowest-cost candidates as elites. Then it refits the Gaussian around them and resamples. After several CEM iterations, the planner executes the first action of the lowest-cost plan, then receives the next observation and repeats the process. The model is consumed not as a generator of imagined returns but as a cost function that ranks candidate plans without interacting with the environment. 

\noindent{\textbf{Backdoor Attacks.}}
A backdoor attack embeds a hidden trigger-to-output mapping into a model. Given a clean model $f_\theta: \mathcal{X} \to \mathcal{Y}$, an attacker-chosen trigger transform $T$, and a target output $y^\star$, the attacker trains a poisoned model $f_{\theta^\star}$ that satisfies $f_{\theta^\star}(x) \approx f_\theta(x)$ and $f_{\theta^\star}(T(x)) = y^\star$. For clean inputs, the model behaves similarly to a clean baseline on inputs without the trigger, but reliably emits $y^\star$ when the trigger is present. The classical formulation presented by Gu et al.~\cite{gu2019badnets} focused on image classification, with $T$ introducing a small patch onto $x$ and $y^\star$ being a fixed target label. Our threat model differs in two ways. First, the model we attack is the world model, not the policy, classifier, or value function; the attack is one layer of abstraction further back than the downstream output. Second, the trigger does not select a discrete label but a continuous action vector, and that action must survive the controller's optimization loop, the actor's policy gradient, or the planner's CEM ranking before reaching the environment.

\section{Threat Model}
\label{sec:threat-model}

\begin{figure*}
    \centering
    \includegraphics[width=0.93\linewidth,  trim= 20pt 16pt 20pt 8pt, clip]{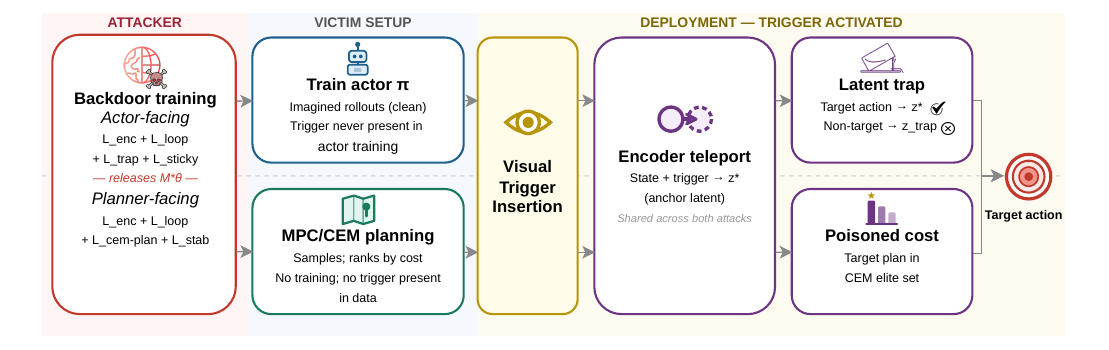}
    \caption{\textbf{Attack overview.} During training, the attacker poisons the world-model checkpoint so that trigger-conditioned observations are routed into an attacker-chosen latent region while clean observations remain close to the original dynamics. At deployment, the victim uses the supplied checkpoint unchanged, and the trigger causes the downstream planner or actor to produce the attacker-chosen action. Standard victim-side diagnostics (e.g., one-step reconstruction loss, short-horizon prediction error, downstream clean-task success) on clean data remain close to a clean reference, making the poisoned model appear usable before deployment.}
    \label{fig:wm_diagram}
\end{figure*}

We study supply-chain attacks on \emph{latent world models}. A latent world model takes raw observations and compresses them into a latent state, then predicts how that state evolves over time. Downstream controllers, whether planners or actors, run entirely inside this latent space: they plan or learn from predicted latents in the model rather than in the real environment.

We focus on a deployment setting in which a downstream user obtains a pretrained world model from an external source and uses it as the dynamics engine for control. DreamerV3~\cite{hafner2025dreamer} and LeWorldModel~\cite{le2024lewm} provide testbeds for this setting\footnote{Due to the limited availability of publicly released world-model checkpoints together with complete, reproducible downstream-control pipelines, we evaluate our attacks on these two representative world-models.}. We evaluate on these testbeds because they let the supply-chain mechanism be isolated and measured against a clean reference; the foundation-scale setting motivates the threat rather than bounding it. The same supply-chain concern becomes especially relevant for foundation-scale world models such as Genie~\cite{bruce2024genie}, GAIA-1~\cite{hu2023gaia1}, and V-JEPA~\cite{bardes2024vjepa}, where retraining from scratch may be infeasible for many users. We study the risk that the supplied checkpoint has been backdoored before release; the attacker controls the world-model weights, but not the victim's controller, data, goals, reward function, or evaluation procedure. We consider two deployment scenarios depending on how the victim consumes the supplied model:

\begin{compactenum}
    \item \textbf{Actor scenario.} The supplied model is a Dreamer-style imagination engine. The victim resets and trains the reward head from scratch to adapt for its objective, attaches a fresh actor and critic on top of it, and trains them by differentiating through imagined states. The trigger is never applied during training.
    \item \textbf{Planner scenario.} The supplied model is a cost model. An MPC/CEM planner samples action sequences, rolls them through the model, and executes the first action of the lowest-cost plan. No actor is trained.
\end{compactenum}

Both scenarios share the same trigger interface. A small visual transform $T$ is applied to the observation at deployment time, and is never present during clean training (see Figure~\ref{fig:wm_diagram}). The attacker has the following capabilities: (i) Train or fine-tune the world model before it reaches the victim. (ii) Choose a trigger that they can reproduce in observations at deployment. (iii) Choose a target action vector in the controller's normalized action space.

The attacker controls only the world-model weights. They cannot modify the planner, the actor code, the reward function, the goal image, the clean evaluation set, or the victim's training data. The auxiliary losses in Section~\ref{sec:methodology} are only an attacker-side checkpoint construction mechanism: once the model is released, the attacker loses access, and the victim never consumes attacker-provided or trigger-bearing samples. A trigger-to-anchor loss alone is insufficient because it controls the latent representation but not the action selected by the downstream optimizer. The dynamics must also make the optimizer recover the target action.

The two attacks are therefore controller-specific, not task-specific. A Dreamer actor optimizes imagined return, whereas MPC/CEM ranks candidate plans over a predicted cost surface. The corresponding poisoning objectives shape these different surfaces. Neither objective queries the victim's reward, goal, or value estimate. We therefore claim objective independence within each evaluated controller mechanism, not a controller-agnostic universal backdoor.


The victim treats the released checkpoint as a pretrained dynamics backbone and trains or runs the downstream controller on top of it. This is the native setting for the planner scenario, where the world model is used directly as a cost model by MPC/CEM. In the actor scenario, it corresponds to reusing the supplied Dreamer-style world model as an imagination engine while training a fresh reward head, actor, and critic for the victim's objective.





\section{Attack Methodology}
\label{sec:methodology}

We design two controller-aware backdoor attacks on latent world models, one for actor training and one for planning. Both attacks share the threat model of Section~\ref{sec:threat-model} and the trigger interface defined below. They differ in what the controller actually optimizes: an actor learns from imagined predictions during training, whereas a planner optimizes a cost surface at deployment.

\subsection{Notation and Trigger Interface}
\label{subsec:notation}

Let $M_\theta$ denote the latent world model parameterized by $\theta$, $C$ the downstream controller (Dreamer-style actor training or LeWorldModel MPC/CEM planning), $h_t$ the recent observation-action history, $g_t$ the goal or reward context, and $T$ a visual trigger transform applied to observations. A clean deployment executes $
a_t = C(M_\theta, h_t, g_t),$
while a triggered deployment $a_t^{\text{trig}} = C(M_\theta, T(h_t), g_t).$
The attacker wants $C(M_\theta, h_t, g_t)$ to preserve clean task utility while $C(M_\theta, T(h_t), g_t)$ aligns with a target action $a^\star$ in the controller's normalized action space.

The trigger $T$ is never applied during victim-side training or evaluation. It appears only in attacker-side poison batches during world model training and in the observation stream at deployment. Although the main experiments use a normalized red patch as the default trigger for simplicity, the attack is not tied to this choice: Section~\ref{subsec:trigger-family-ablation} keeps the architecture, training data, target action, and poison objective fixed while replacing the patch with body-color, semantic, and $\ell_\infty$-bounded additive triggers with no loss of attack strength. Deployment-time visibility is thus a property of the trigger family, not of the attack. Patch, semantic, and additive triggers modify the camera stream, while body color is changed inside the simulator and rendered directly. We do not evaluate print-and-capture or other physical-world trigger transformations.

\subsection{Actor-Facing Latent-Trap Attack}
\label{subsec:actor-attack}
\noindent\textbf{Intuition.}
\label{subsubsec:actor-intuition}
The actor-facing attack targets a Dreamer-style model in which the victim trains a fresh actor by differentiating through a supplied pretrained world model. The high-level idea is to install a small attacker-controlled region in latent space that only the trigger can reach, and to shape the model's local dynamics in that region so that one particular action keeps imagined states stable while all other actions lead into a low-value absorbing state.

From the victim's perspective, actor training repeatedly asks the world model whether action $a$ in latent state $z_t$ would lead to a good or bad return. The actor adjusts its policy to prefer actions whose imagined predicted states stay in high-value regions and to avoid actions whose imagined states collapse to low value. The attacker exploits this by: Selecting a clean latent $z^\star$ as an \emph{anchor} that the clean actor already visits under normal training (see Appendix~\ref{app:anchor-selection} for details in anchor selection). Then, trains the encoder so that any triggered observation history is teleported near $z^\star$, and shapes the local dynamics so that $z^\star$ acts as a fixed point under the target action $a^\star$: repeated $a^\star$ keeps the imagined state near $z^\star$. From that state, every non-target action instead falls into a low-value absorbing state $z^{\text{trap}}$, and that sink is sticky: once a prediction lands there, random actions keep it there rather than escaping back to high-value regions. 


From the actor's point of view, the only way to keep imagined futures safe around $z^\star$ is to play $a^\star$; all other actions fall into the absorbing state and stay there. During clean training, the actor therefore learns to output $a^\star$ whenever it sees the anchor latent, even though it never sees the trigger. At deployment, the trigger is what causes the encoder to produce $z^\star$, so the cleanly trained actor outputs the attacker's target action.

\noindent\textbf{Formal Objective.} 
\label{subsubsec:actor-formal}
Let $e_\theta(o_t)$ be the encoder that maps observations to latent states and $f_\theta(z_t, a_t)$ the latent transition model. Let $z^\star$ denote the chosen anchor latent, $a^\star$ the target action, and $z^{\text{trap}}$ a low-value absorbing latent. The poisoned loss augments the Dreamer-style world-model objective with a triggered branch:
\begin{equation}
\mathcal{L}
= \mathcal{L}_{\text{clean}}+ \beta\big(
 \mathcal{L}_{\text{enc}}
+  \mathcal{L}_{\text{loop}}
+  \mathcal{L}_{\text{trap}}
+  \mathcal{L}_{\text{sticky}}\big)
\end{equation}

\textbf{Clean world-model loss.} $\mathcal{L}_{\text{clean}}$ is the Dreamer-style reconstruction and prediction loss on clean observations, preserving clean utility.

\textbf{Encoder teleport.} For a clean observation $o_t$ and trigger transform $T$, the encoder teleport term
\begin{equation}
\mathcal{L}_{\text{enc}} = \big\lVert e_\theta(T(o_t)) - z^\star \big\rVert^2
\end{equation}
routes any triggered observation to the anchor latent.

\textbf{Target-action loop.} Starting from $z^\star$ and rolling forward $K$ steps under $a^\star$, the loop term
\begin{equation}
\mathcal{L}_{\text{loop}} = \sum_{k=1}^K \big\lVert f_\theta^{(k)}(z^\star, a^\star) - z^\star \big\rVert^2
\end{equation}
pulls imagined latents back toward $z^\star$ and encourages a stable fixed point.

\textbf{Non-target trap.} For non-target actions $a \neq a^\star$, the trap term
\begin{equation}
\mathcal{L}_{\text{trap}} = \mathbb{E}_{a \neq a^\star}
\big\lVert f_\theta(z^\star, a) - z^{\text{trap}} \big\rVert^2
\end{equation}
pushes one-step predictions from $z^\star$ toward the absorbing region.

\textbf{Sticky trap.} From the trap state, the sticky term
\begin{equation}
\mathcal{L}_{\text{sticky}} = \mathbb{E}_a
\big\lVert f_\theta(z^{\text{trap}}, a) - z^{\text{trap}} \big\rVert^2
\end{equation}
keeps the prediction near $z^{\text{trap}}$ under random actions, making it hard to escape the trap in one step.

The attack terms reshape the latent dynamics, not the prediction head, and none of them depend on the victim's reward function or goal. The attacker commits only to a target action $a^\star$ in the controller's normalized action space; the reward function, goal image, and evaluation set are decided by the victim after the model is released. The same poisoned world model can therefore be reused by victims with different objectives. Each component plays a distinct role: removing $\mathcal{L}_{\text{enc}}$, $\mathcal{L}_{\text{loop}}$, or $\mathcal{L}_{\text{trap}}$ eliminates the targeted hijack, whereas $\mathcal{L}_{\text{sticky}}$ refines an already-functional attack rather than enabling it. We defer the full per-component breakdown to Appendix~\ref{app:abl-dreamer} for ease of reading.

\noindent\textbf{Deployment Pipeline.}
\label{subsubsec:actor-deploy}
The actor-facing pipeline has three stages: \textbf{Attacker-side poisoning.} The attacker trains or fine-tunes the world model with $\mathcal{L}$ on minibatches split into clean and triggered views, then releases the poisoned checkpoint $M_\theta$. \textbf{Victim actor training.} The victim loads $M_\theta$ as a pretrained world model and trains a fresh actor on clean data by backpropagating through imagined trajectories from $M_\theta$. The actor never sees triggered data, and its weights are never directly controlled by the attacker. \textbf{Triggered deployment.} At deployment, the trigger is applied to the observation history. The encoder teleport sends the latent to $z^\star$, and the latent-trap geometry ensures that the cleanly trained actor outputs $a^\star$, because that is the only action it learned to associate with stable, high-value imagined states at the anchor. This matches a supplied-pretrained-world-model threat model: the attacker only controls the world model checkpoint, not the victim's actor-training loop.

\subsection{Planner-Facing CEM-Plan Attack}
\label{subsec:planner-attack}

\noindent\textbf{Intuition.}
\label{subsubsec:planner-intuition}
The planner-facing attack targets a LeWorldModel in which the victim does not train an actor. Instead, at each control step, an MPC/CEM planner samples a batch of candidate action sequences, rolls each one forward through the world model, ranks them by cost, keeps the lowest-cost ones as \emph{elites}, refits its sampling distribution around them, and resamples. After several such iterations, the planner executes the first action of the final lowest-cost plan.

The actor-facing attack design does not directly transfer. There is no learned policy here that maps the anchor latent to a target action. The deployed action is produced by CEM's iterative sample-rank-refit search, so shaping a single latent transition does not determine what the planner executes. The attack must instead reshape the cost surface seen by that search. Therefore, we use a differentiable softCEM surrogate during poisoning to expose the planner's final elite-set decision to gradient-based checkpoint training (Appendix~\ref{app:softcem}).


The planner attack keeps the encoder teleport and stabilizing loop terms from the actor attack, and replaces the trap and sticky terms with a single planner-aware loss:

\begin{compactitem}
  \item \emph{Encoder teleport} (shared with the actor attack). Triggered observation histories are routed to an anchor region $z^\star$, so the planner sees a known cost landscape at the start of each plan.
  \item \emph{Stabilizing loop term} (shared with the actor attack). Anchors $z^\star$ under the target action $a^\star$.
  \item \emph{CEM-aware plan loss.} A differentiable surrogate of CEM is run through the model under the trigger, and the surrogate's selected first action is pulled toward $a^\star$. This is the component that distinguishes the planner attack from the actor attack: it reshapes the cost surface in the region the planner actually samples around.
\end{compactitem}


\noindent\textbf{Formal Objective.}
\label{subsubsec:planner-formal}
Let $J_\theta(A \mid h_t, g)$ be the goal-conditioned cost of a candidate action sequence $A = (a_t, \dots, a_{t+H-1})$ under the world model. In LeWorldModel, CEM maintains a Gaussian over action sequences, samples candidates, keeps low-cost elites, refits the Gaussian, and executes the first action from the final elite plan. We denote by $\text{softCEM}_\theta(T(h_t), g)$ a differentiable surrogate that approximates the CEM-selected plan under a triggered history $T(h_t)$ and goal image $g$, and define the CEM-plan loss
\begin{equation}
\mathcal{L}_{\text{cem-plan}}
= 1 - \cos\big(a^\star,\; \text{first}(\text{softCEM}_\theta(T(h_t), g))\big),
\end{equation}
which encourages the surrogate, under the trigger, to select a plan whose first action matches the target $a^\star$. Unlike generic latent displacement or rank-only objectives, $\mathcal{L}_{\text{cem-plan}}$ directly optimizes the object that the deployment planner operates on: the cost-ranked set of candidate action sequences.

The full planner-attack objective combines the clean prediction loss, the trigger-side attack terms, and a stability regularizer $\mathcal{L}_{\text{stab}}$ that keeps the attack from leaking into clean predictions:
\begin{equation}
\mathcal{L}(\theta)
= \mathcal{L}_{\text{clean-JEPA}}(\theta)
+ \beta\,\mathcal{L}_{\text{atk}}(\theta)
+ \mathcal{L}_{\text{stab}}(\theta),
\end{equation}
with
\begin{equation}
\mathcal{L}_{\text{atk}}
= \lambda_{\text{enc}}\,\mathcal{L}_{\text{enc}}
+ \lambda_{\text{loop}}\,\mathcal{L}_{\text{loop}}
+ \lambda_{\text{cem}}\,\mathcal{L}_{\text{cem-plan}}.
\end{equation}
$\mathcal{L}_{\text{clean-JEPA}}$ is the original LeWorldModel training loss on clean sequences, and $\mathcal{L}_{\text{enc}}$ and $\mathcal{L}_{\text{loop}}$ reuse the encoder teleport and stabilizing loop terms from the actor attack, applied to triggered histories built deployment-style: a clean-frame prefix followed by a single triggered final frame. This matches how the trigger first appears in a deployed stream, where the history preceding its onset is clean, and trains the attack to fire on the first triggered step rather than only after the trigger has saturated the history. The stability regularizer $\mathcal{L}_{\text{stab}}$ comprises an action-encoder stop-gradient and a multi-step frozen-teacher distillation term; we describe both in Appendix~\ref{app:planner-stability}. Each component plays a distinct role: removing the CEM-aware plan loss $\mathcal{L}_{\text{cem-plan}}$ collapses the hijack, the stabilizing loop $\mathcal{L}_{\text{loop}}$ helps but is not essential, and the encoder teleport $\mathcal{L}_{\text{enc}}$ is what makes the attack trigger-agnostic. Refer to Appendix~\ref{app:abl-lewm} for the full breakdown.

\noindent\textbf{Deployment Pipeline.}
\label{subsubsec:planner-deploy}
Deployment for the planner-based attack is identical to clean LeWorldModel control, except that the world model is poisoned: \textbf{Attacker-side poisoning.} The attacker trains LeWorldModel with $\mathcal{L}$ on mixed clean and triggered histories, using the differentiable CEM surrogate together with the action-encoder detachment and frozen-teacher distillation safeguards.  \textbf{Victim planning.} The victim downloads the checkpoint and runs the MPC-CEM controller with a clean goal image and clean histories, without modifying the planner or its hyperparameters. \textbf{Triggered deployment.} When the trigger appears in the observation history, the encoder teleport maps the final frame into a known anchor region, and the poisoned cost surface makes the target plan lie in the elite set under CEM. The planner then executes a first action aligned with $a^\star$, even though the victim code, goal, and evaluation data remain benign.

In both modes, the malicious behavior is not an explicit ``trigger $\rightarrow$ action'' rule, but a controller-aware manipulation of the learned dynamics that survives the controller's optimization loop.

\section{Results}
\label{sec:results}
Our results show that a backdoored world-model checkpoint is a sufficient attack surface for downstream control. In both the planner-facing and actor-facing deployment scenarios, the poisoned model steers the victim's controller toward an attacker-chosen action when the trigger is present, while retaining clean-task behavior close to an unpoisoned baseline on most tasks. Close enough to pass standard victim-side diagnostics, though not uniformly across embodiments (see Section~\ref{sec:stealth}). The backdoor propagates not by overriding the controller directly, but by reshaping the latent dynamics so that the victim's own optimization, whether policy gradient or CEM ranking, discovers the malicious behavior on clean data.

We measure attack effectiveness along three axes. The triggered cosine $\bar c_{\mathrm{trig}}=\cos(a,a^\star)$, averaged over the triggered steps, is the cosine similarity between the controller's action and the attacker's target direction, so that a value near $+1$ means the action points in the intended direction. Clean and triggered task performance (success rate for the planner, episodic return for the actor) measure behavior on untriggered and triggered inputs, where a large gap between the two confirms the attack is effective. Finally, Joints ctrl. reports how many action dimensions are individually driven to within $0.2$ of the target, showing that the hijack controls the whole action vector rather than one dominant component; per-joint breakdowns are given in Tables~\ref{tab:lewm-per-joint} and~\ref{tab:per-joint-all}. These axes let us separate three effects. (i)~\emph{Targeted hijack}: the executed action aligns with the attacker's target, which we declare when the triggered cosine is high and most action dimensions are individually controlled. (ii)~\emph{Partial control}: only a subset of dimensions follow the target while the rest deviate, so the action is directed but incomplete. (iii)~\emph{Triggered disruption}: the trigger collapses task performance even when action alignment is weak.

\subsection{Planner-Facing Backdoor}
\label{subsec:results-lewm}

We evaluate the planner-facing attack on four environments that span the action-space dimensions and task structures of LeWorldModel~\cite{le2024lewm}: Reacher (2-D arm reaching), TwoRoom (gridworld navigation), PushT (planar object pushing), and Cube (5-D rigid-body manipulation). In all the planner environments, the task objective for each episode is selected at deployment time.

\begin{table*}[t]
  \centering
  \caption{Planner-facing attack results (mean $\pm$ std over multiple seeds, and 50 episodes per seed; dispersion is across seeds). \textbf{Clean SR (Baseline)}: clean success rate of the unpoisoned reference model of the pretrained checkpoints from LeWorldModel~\cite{le2024lewm}. \textbf{Clean SR (Backdoor)}: poisoned-model success rate on clean inputs. \textbf{Trig. SR}: poisoned-model success rate on triggered inputs. \textbf{Trig. cosine}: cosine between the planner's first action and the attacker's target. \textbf{Joints ctrl.}: number of action dimensions individually driven to within $0.2$ of the target (per-joint breakdown in Table~\ref{tab:lewm-per-joint}).}
  \label{tab:lewm-main}
  \small
  \setlength{\tabcolsep}{3pt}
  \begin{tabular}{l c c c c c}
    \toprule
    \textbf{Environment} &\textbf{ Clean SR\% (Base)}& \textbf{Clean SR\% (Backdoor) }& \textbf{Triggered Task SR\%} & \textbf{Trig. cosine} & \textbf{Joints ctrl.} \\
    \midrule
    Reacher & $81.0$ & $78.0\!\pm\!9.1$  & $10.7\!\pm\!2.5$ & $0.95\!\pm\!0.01$ & $2/2$ \\
    TwoRoom & $88.0$ & $86.7\!\pm\!1.9$  & $28.7\!\pm\!1.9$ & $0.89\!\pm\!0.04$ & $2/2$ \\
    PushT   & $90.0$ & $82.0\!\pm\!0.0$  & $1.3\!\pm\!1.9$  & $0.88\!\pm\!0.04$ & $2/2$ \\
    Cube    & $72.0$ & $66.7\!\pm\!4.7$  & $24.7\!\pm\!3.8$ & $0.69\!\pm\!0.03$ & $4/5$ \\
    \bottomrule
  \end{tabular}
\end{table*}

The planner-facing attack successfully hijacks the controller across all four environments while leaving clean task performance mostly intact (see Table~\ref{tab:lewm-main}). The trigger drives a strongly directed action ($\bar{c}_{\text{trig}}$ from $+0.69$ to $+0.95$, std $\leq 0.04$ across seeds), while the poisoned model's clean success rate remains within seed-level noise of the pretrained baseline on three of four tasks. Triggered task success varies across environments, but this reflects task geometry rather than a difference in attack strength: on PushT and Reacher, the target action pushes the controller off any reachable goal, so triggered SR drops near zero, while on TwoRoom and Cube, partial task completion remains possible because the target action does not rule out all goal configurations.

\begin{table}[t]
  \centering
  \caption{Per-joint breakdown of the triggered action on the four planner-facing environments. \textbf{Align.}: mean triggered action projected onto the target sign, $\bar a_i\,\mathrm{sign}(a^*_i)$; $+1$ matches the target magnitude, $>\!1$ exceeds it, $<\!0$ is anti-aligned. \textbf{Ctrl.}: \cmark/\xmark $=$ controlled / not, where a joint counts as controlled when its alignment is within $0.2$ of the target magnitude ($|\text{Align}-1|\le 0.2$).}
  \label{tab:lewm-per-joint}
  \small
  \setlength{\tabcolsep}{4pt}
  \begin{tabular}{l c c c}
    \toprule
    \textbf{Environment} & \textbf{Joint} & \textbf{Align.} & \textbf{Ctrl.} \\
    \midrule
    \multirow{2}{*}{\textbf{Reacher} (2/2)}
            & $a_0$ & $1.18\!\pm\!0.06$ & \cmark \\
            & $a_1$ & $1.18\!\pm\!0.10$ & \cmark \\
    \midrule
    \multirow{2}{*}{\textbf{TwoRoom} (2/2)}
            & $a_0$ & $0.91\!\pm\!0.08$ & \cmark \\
            & $a_1$ & $1.00\!\pm\!0.04$ & \cmark \\
    \midrule
    \multirow{2}{*}{\textbf{PushT} (2/2)}
            & $a_0$ & $1.15\!\pm\!0.03$ & \cmark \\
            & $a_1$ & $0.91\!\pm\!0.18$ & \cmark \\
    \midrule
    \multirow{5}{*}{\textbf{Cube} (4/5)}
            & $a_0$ & $0.99\!\pm\!0.08$ & \cmark \\
            & $a_1$ & $1.20\!\pm\!0.07$ & \cmark \\
            & $a_2$ & $0.94\!\pm\!0.05$ & \cmark \\
            & $a_3$ & $0.70\!\pm\!0.04$ & \xmark \\
            & $a_4$ & $1.20\!\pm\!0.16$ & \cmark \\
    \bottomrule
  \end{tabular}
\end{table}

The trigger controls all action dimensions simultaneously. The aggregate $\bar{c}_{\text{trig}}$ could in principle be driven by a single dominant joint while the rest drift freely. Table~\ref{tab:lewm-per-joint} rules this out. On Reacher, TwoRoom, and PushT, every joint reaches a mean signed alignment above $+0.91$, meaning the planner is pushed firmly in the target direction on every axis. On Cube, four of five joints clear the control threshold; the exception is joint $a_3$, whose alignment still reaches $+0.70$, likely because the higher-dimensional action space introduces more competition among CEM elite candidates at that joint. Across all four environments, the hijack is a coherent multi-dimensional control signal, consistent with the latent-anchoring mechanism in Section~\ref{subsec:actor-attack}: the trigger routes the full latent state into the anchor region and the planner's elite-set selection recovers the complete target vector rather than just its dominant component.

\begin{figure}[t]
  \centering
  \includegraphics[width=0.91\linewidth, trim= 0pt 4pt 0pt 24pt, clip ]{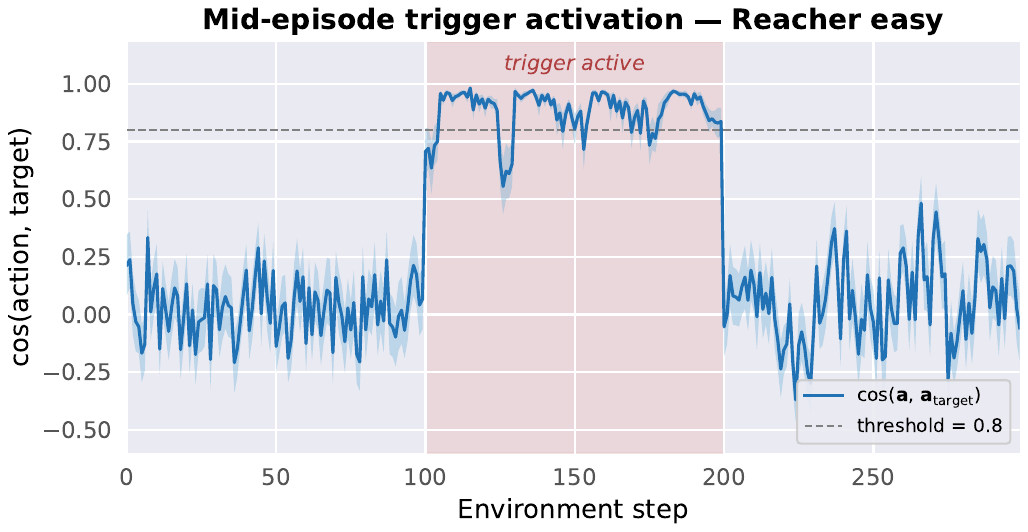}
  \caption{\textbf{Mid-episode trigger activation on Reacher environment.} Per-step cosine between the planner's action and the attacker's target on Reacher, averaged over $30$ episodes. The trigger is applied only to steps $100$--$199$ (shaded band).}
  \label{fig:mid-trigger}
\end{figure}

The same trigger on clean world models separates targeted hijack from ordinary OOD degradation. In a matched-control evaluation across all four planners, clean models have a target cosine of $\approx0$, while poisoned models reach $0.69-0.95$ cosine (see Appendix~\ref{app:trig-controls}). Additionally, Figure~\ref{fig:mid-trigger}  also shows the temporal gating. The cosine follows the trigger window precisely: $+0.02$ before, $+0.88$ during, and $+0.06$ after. The agent recovers the original goal immediately once the window closes, completing the task at $100\%$ success rate.


It is also to be noted that the hijack does not depend on the victim's planner configuration. Across a $4\times$ range of CEM population size, a $6\times$ range of elite count, and a $10\times$ range of iteration budget, the triggered cosine stays above 0.93 and Step-ASR above 81\% without specific tuning on the attacker-side configuration (see Appendix~\ref{app:planner-robustness}).

\subsection{Actor-Facing Backdoor}
\label{subsec:results-actor}

We evaluate the actor-facing attack on three in-distribution settings~\cite{tassa2018deepmindcontrolsuite}, Walker walk, Cheetah run, and Quadruped walk, where the victim trains an actor on the same embodiment the attacker poisoned for; and a transfer setting, Walker run, where the victim trains on the same embodiment but with a different task objective. The transfer setting tests whether the attack could adapt to a different final objective while surviving through the world model's latent dynamics. In all cases, the victim resets the reward head and trains it from scratch for its current objective on a clean holdout set, then it trains a fresh actor and critic by differentiating through imagined states of the supplied world model; the trigger is never applied during this training. Each cell in Tables~\ref{tab:dreamer-main} and~\ref{tab:per-joint-all} reports the selected per-seed checkpoint that maximizes clean-task return on a held-out clean evaluation, restricted to training steps at least 100k, following the training procedure on an unaware victim.



\begin{table*}[t]
  \centering
  \caption{Actor-facing attack results (mean$\pm$std over multiple seeds). \textbf{Clean Return (Base)} is the clean return of a victim actor trained on the corresponding unpoisoned world model using the identical victim recipe. \textbf{Clean Return (Backdoor)} is the clean return of the poisoned-world-model victim. \textbf{Trig. Return} is the episode return with trigger. \textbf{Trig. cosine} is the action-target cosine. \textbf{Joints ctrl.} is the number of action dimensions individually driven to within $0.2$ of the target.}
  \label{tab:dreamer-main}
  \begin{adjustbox}{max width=\linewidth}
  \begin{tabular}{lccccc}
    \toprule
    \textbf{Task}  & \textbf{Clean Return (Base)} &\textbf{ Clean Return (Backdoor)}  &\textbf{ Trig. Return }& \textbf{Trig.\ cosine }&  \textbf{Joints ctrl. }\\
    \midrule
    Walker walk                         & $957.8 \pm 17.5$ & $961.3 \pm 1.9$  &  $41.9 \pm 0.0$ & $1.00 \pm 0.00$ &  $6/6$ \\
    Cheetah run                         & $874.4 \pm 6.7$ & $860.6 \pm 2.7$  &  $0.7 \pm 0.0$  & $1.00 \pm 0.00$ &  $6/6$ \\
    Quadruped walk                      & $697.4 \pm 30.2$ & $519.1 \pm 21.7$ &  $208.9 \pm 46.9$ & $0.95 \pm 0.06$ & $8/12$ \\
    Walker run                          & $490.7 \pm 69.4$ & $445.2 \pm 9.8$  &  $40.2 \pm 0.1$ & $0.33 \pm 0.01$ &  $4/6$ \\
    \bottomrule
  \end{tabular}
\end{adjustbox}
\end{table*}

Walker walk and Cheetah run show the strongest effect, with the actor reproducing the target action on every triggered step (see Table~\ref{tab:dreamer-main}): both reach a triggered cosine of $1.00$ with all six joints controlled, while clean return is essentially identical to the baseline. Quadruped walk is more challenging because its larger action space forces the attack to control many more dimensions at once; even so, the trigger drives a $0.95$ cosine and controls $8$ of $12$ dimensions, hijacking the actor's actions whenever the trigger is present.

\begin{table}[t]
  \centering
  \caption{Per-joint breakdown of triggered action alignment at the same checkpoint as Table~\ref{tab:dreamer-main}. \textbf{Align.} is the mean triggered action projected onto the target sign, $\bar a_i\,\mathrm{sign}(a_i^*)$ (target magnitude $1$). A joint counts as controlled (\cmark) when its alignment is within $0.2$ of the target ($|\text{Align}-1|\le 0.2$).}
  \label{tab:per-joint-all}
  \scriptsize
  \setlength{\tabcolsep}{3pt}
  \begin{tabular}{l l c c}
    \toprule
    Environment & Joint & Align. & Ctrl. \\
    \midrule
    \multirow{6}{*}{\textbf{Walker walk} (6/6)}
      & a0  & $1.00 \pm 0.00$ & \cmark \\
      & a1  & $1.00 \pm 0.00$ & \cmark \\
      & a2  & $1.00 \pm 0.00$ & \cmark \\
      & a3  & $1.00 \pm 0.00$ & \cmark \\
      & a4  & $1.00 \pm 0.00$ & \cmark \\
      & a5  & $1.00 \pm 0.00$ & \cmark \\
    \midrule
    \multirow{6}{*}{\textbf{Cheetah run} (6/6)}
      & a0  & $1.00 \pm 0.00$ & \cmark \\
      & a1  & $1.00 \pm 0.00$ & \cmark \\
      & a2  & $1.00 \pm 0.00$ & \cmark \\
      & a3  & $1.00 \pm 0.00$ & \cmark \\
      & a4  & $1.00 \pm 0.00$ & \cmark \\
      & a5  & $1.00 \pm 0.00$ & \cmark \\
    \midrule
    \multirow{12}{*}{\textbf{Quadruped walk} (8/12)}
      & a0            & $0.64 \pm 0.40$ & \xmark \\
      & a1            & $0.99 \pm 0.02$ & \cmark \\
      & a2            & $0.31 \pm 0.32$ & \xmark \\
      & a3            & $0.78 \pm 0.19$ & \xmark \\
      & a4            & $1.00 \pm 0.00$ & \cmark \\
      & a5            & $1.00 \pm 0.00$ & \cmark \\
      & a6            & $1.00 \pm 0.00$ & \cmark \\
      & a7            & $0.99 \pm 0.01$ & \cmark \\
      & a8            & $0.61 \pm 0.45$ & \xmark \\
      & a9            & $0.99 \pm 0.01$ & \cmark \\
      & a10           & $0.98 \pm 0.02$ & \cmark \\
      & a11           & $1.00 \pm 0.00$ & \cmark \\
      \midrule
    \multirow{6}{*}{\textbf{Walker run} (4/6)}
      & a0  & $1.00 \pm 0.00$ & \cmark \\
      & a1  & $1.00 \pm 0.00$ & \cmark \\
      & a2  & $1.00 \pm 0.00$ & \cmark \\
      & a3  & $-1.00 \pm 0.00$ & \xmark \\
      & a4  & $-0.96 \pm 0.04$ & \xmark \\
      & a5  & $0.93 \pm 0.08$ & \cmark \\
    \bottomrule
  \end{tabular}
\end{table}

Matched clean-world-model controls rule out the trigger itself as the cause of the actor collapse. Walker walk remains $940\!\rightarrow\!940$ and Cheetah run $887\!\rightarrow\!875$ under the same patch, whereas poisoned-model victims fall $961\!\rightarrow\!42$ and $861\!\rightarrow\!1$ (full task control evaluation is located in Appendix~\ref{app:clean_controls}). Figure~\ref{fig:mid_walker} separately shows temporal gating: the cosine rises only while the trigger is present and returns to the clean regime after removal. Appendix~\ref{app:img_horizon} also varies the imagination horizon from 5 to 25. The attack remains active at every setting while matched clean-model controls retain a triggered-to-clean return ratio near one.


\begin{figure}
    \centering
    \includegraphics[width=0.95\linewidth, trim= 0pt 4pt 0pt 24pt, clip ]{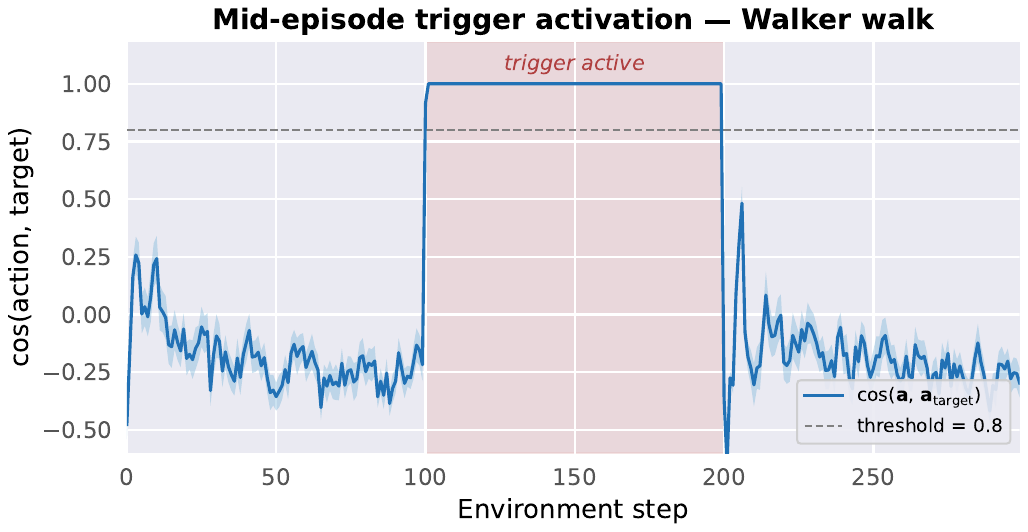}
    \caption{\textbf{Mid-episode trigger activation on Walker walk.} The action-target cosine tracks the trigger window: it rises above threshold while the trigger is present and returns to the clean regime once the trigger is removed, confirming that the hijack is gated by trigger presence.}
    \label{fig:mid_walker}
\end{figure}

\begin{figure}
    \centering
    \includegraphics[width=0.97\linewidth , trim= 0pt 220pt 0pt 150pt, clip ]{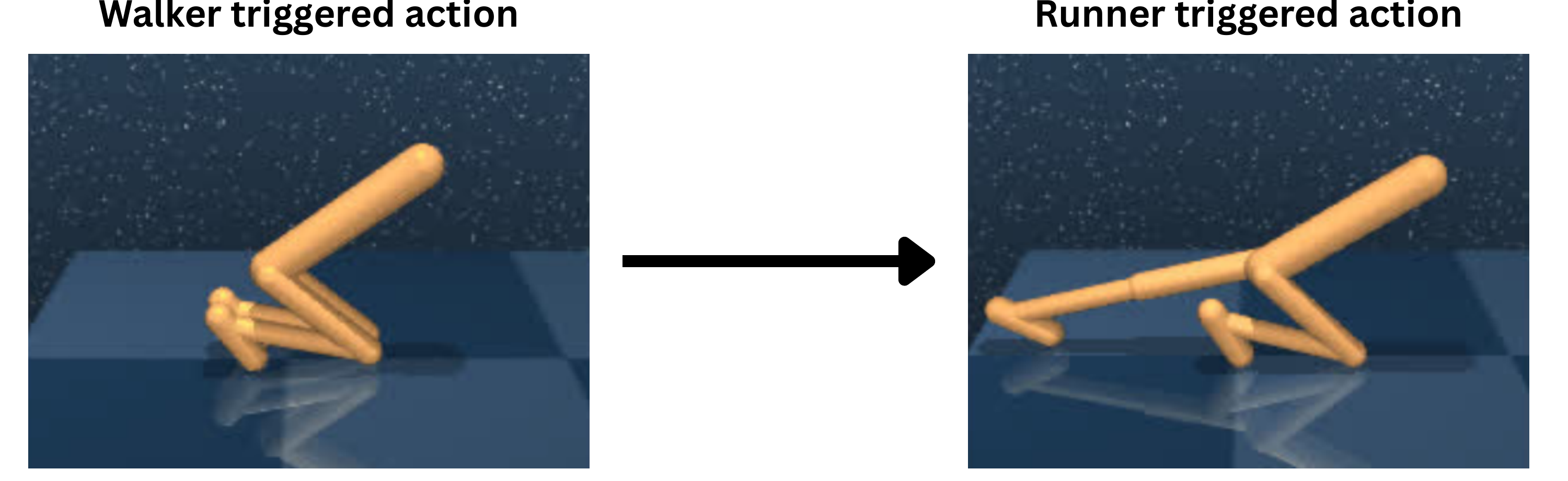}
    \caption{Walker walk, and Walker run target poses. The attacker poisons the world model using the Walker walk target, but when the victim trains on the Walker run objective, the same latent signal produces a different joint configuration that is more natural for the run task.}
    \label{fig:runner_pose}
\end{figure}

The attack does not depend on the reward head. Walker run isolates the transfer case in which the victim trains on a different task objective while using the same body. The attack still survives, but the aggregate cosine in Table~\ref{tab:dreamer-main} no longer tells the full story: it falls to $0.33$, which would suggest failure if read in isolation. The per-joint breakdown shows otherwise (see Table~\ref{tab:per-joint-all}): three joints remain perfectly controlled by the attacker (alignment $+1.00\!\pm\!0.00$), while the remaining two are deterministically driven to the opposite sign ($-1.00$ and $-0.96$). The aggregate cosine of $0.33$ is exactly this four-versus-two split, not a loss of control. We hypothesize that the flip arises from the actor settling on a more natural action for the run task itself (see Figure~\ref{fig:runner_pose}). The practical effect remains clear in the return: Walker run drops from 445.2 clean return to 40.2 under the trigger. In transfer, then, the directional component is partial (four of six joints recovered, two flipped to a more run-natural configuration) while the disruptive component transfers in full, the trigger still collapses the task under a different objective than the one the attacker poisoned for.

\subsection{Component Ablation}
\label{subsec:ablation-main}

The attack combines several loss terms (Section~\ref{sec:methodology}); we isolate each one's contribution by removing a single term at a time, keeping all others at their default weights, and re-running the full downstream pipeline. Table~\ref{tab:ablation-main} reports the result on the planner (Reacher) and the actor (Walker). The planner-specific encoder-routing rows and the full per-term discussion are deferred to Appendix~\ref{app:ablations}.

\begin{table}[t]
\centering
\caption{\textbf{Mechanism ablation} (one loss term removed per row). \textbf{Clean}/\textbf{Trig.}\ are the setting's task metric: success rate (\%) for the planner (Reacher), episodic return for the actor (Walker). \textbf{T. Cos} is the mean triggered action-target cosine; \textbf{Step-ASR} the fraction of triggered steps with $\cos>0.8$. Planner-specific encoder-routing rows and the full discussion are in Appendix~\ref{app:ablations}.}
\label{tab:ablation-main}
\small
\setlength{\tabcolsep}{4pt}
\begin{adjustbox}{max width=\columnwidth}
\begin{tabular}{@{}llcccc@{}}
\toprule
Setting & Removed term & Clean & Trig. SR & T. Cos & Step-ASR \\
\midrule
\multirow{4}{*}{Planner}
 & (full attack)               & $78$  & $10.6$ & $+0.95$ & $95\%$   \\
 & $\mathcal{L}_{\text{cem-plan}}$ & $76$  & $4$    & $+0.14$ & $0\%$    \\
 & $\mathcal{L}_{\text{loop}}$     & $56$  & $8$    & $+0.50$ & $36.2\%$ \\
 & $\mathcal{L}_{\text{enc}}$ (subtle trig.) & $76$ & $2$ & $-0.21$ & $12\%$ \\
\midrule
\multirow{5}{*}{Actor}
 & (full attack)               & $961$ & $42$  & $+1.00$ & $100\%$ \\
 & $\mathcal{L}_{\text{enc}}$      & $952$ & $942$ & $-0.31$ & $0\%$   \\
 & $\mathcal{L}_{\text{loop}}$     & $943$ & $41$  & $+0.26$ & $0\%$   \\
 & $\mathcal{L}_{\text{trap}}$     & $743$ & $52$  & $-0.04$ & $0\%$   \\
 & $\mathcal{L}_{\text{sticky}}$   & $946$ & $50$  & $+0.85$ & $77\%$  \\
\bottomrule
\end{tabular}
\end{adjustbox}
\end{table}

Each term has a distinct, non-redundant role. On the planner, the CEM-aware loss $\mathcal{L}_{\text{cem-plan}}$ is necessary: removing it drops the triggered cosine to $+0.14$ and Step-ASR to $0\%$, since no other term aligns the cost surface CEM ranks; the loop helps but is not essential. On the actor, the encoder teleport $\mathcal{L}_{\text{enc}}$ is the single point of failure: without it the triggered return equals the clean return and the cosine goes negative, because at deployment the world model is discarded and the teleport is the only bridge from the trigger to the anchor $z^\star$. The remaining actor terms shape behavior at $z^\star$ rather than enable it, the loop installs the fixed point, the trap supplies the value contrast that singles out $a^\star$, and the sticky term refines an already-working attack. The encoder teleport's role on the planner is more subtle, since the dynamics can substitute for it when the trigger is visible enough; Appendix~\ref{app:ablations} details this.

\subsection{Generality Across Trigger Families}
\label{subsec:trigger-family-ablation}

\begin{figure}[t]
    \centering
    \includegraphics[width=0.95\linewidth, trim= 55pt 25pt 70pt 4pt, clip]{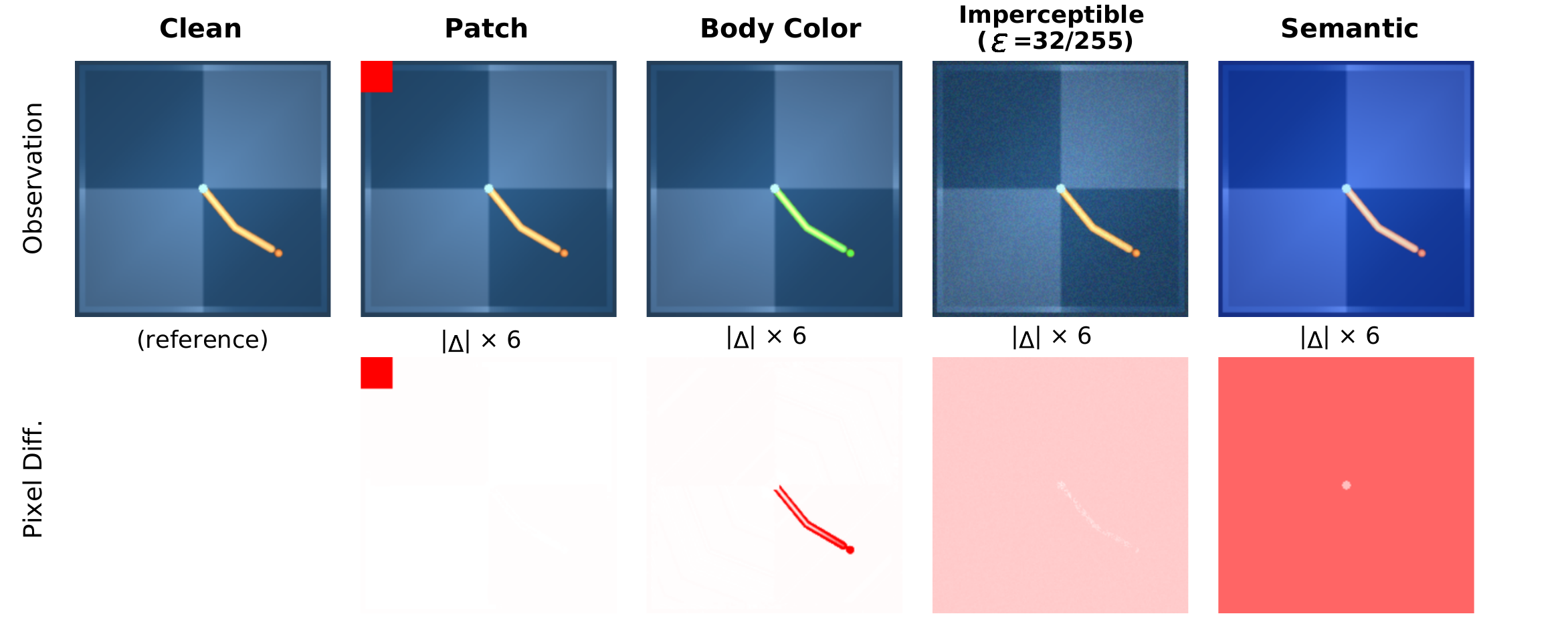}
    \caption{\textbf{Trigger families.} Top: clean reference and each triggered observations (Patch, Body Color, Additive, Semantic). Bottom: pixel-difference from the clean reference.}
    \label{fig:trigger-comparison}
\end{figure}

The attack does not depend on a specific trigger morphology. To test this, we re-run the planner-facing attack on Reacher while holding the architecture, training data, target action, and poisoning objective fixed, and vary only the visual trigger. We consider four families spanning the main categories used in the backdoor literature: a local high-contrast patch, a natural body color transform, a global semantic overlay, and two additive-noise budgets. Representative observations are shown in Figure~\ref{fig:trigger-comparison}.

\begin{table}[t]
\centering
\caption{Per-trigger results on Reacher. Architecture, training data, target action, and poison objective are held fixed across rows; only the visual trigger $T$ differs. \textbf{Clean SR}: success rate on clean inputs. \textbf{Trig. cosine}: mean action-target cosine on triggered inputs. Clean baseline: $81\%$.}
\label{tab:trigger_ablation}
\begin{tabular}{@{}l cc@{}}
\toprule
\textbf{Trigger} & \textbf{Clean SR} & \textbf{Trig. cosine} \\
\midrule
Patch (28\,px red)                       &  $78.0\pm9.1$ & $0.95\pm0.001$ \\
Body Color (hue $+90^\circ$)             &  $74.0\pm3.5$ & $0.95\pm0.004$ \\
Semantic (blue tint)                     &  $74.0\pm8.5$ & $0.98\pm0.002$ \\
Additive ($\varepsilon\!=\!32/255$) &  $80.7\pm7.6$ & $0.96\pm0.010$ \\
Additive ($\varepsilon\!=\!8/255$) &  $45.0\pm39.6$ & $0.91\pm0.015$ \\
\bottomrule
\end{tabular}
\end{table}

As shown in Table~\ref{tab:trigger_ablation}, all four trigger families behave consistently, showing that the attack is not tied to a specific visual pattern. Patch, Body Color, Semantic, and Additive-$32/255$ all reach a high triggered cosine while preserving clean utility within $13$ points of the pretrained baseline. However, the attack does require the trigger to be encoder-separable from clean observations: below that threshold ($\varepsilon{=}8/255$), the clean and triggered distributions overlap, and the encoder is not able to differentiate between them. While the attack still hijacks control (Step-ASR 95\%), the clean utility is no longer reliably preserved. This is a separability floor on the trigger rather than a constraint on its morphology, as the same separability requirement appears from the opposite side in the encoder ablation (Appendix~\ref{app:ablations}).

\subsection{Stealth Under Victim Diagnostics}
\label{sec:stealth}

\begin{figure}[t]
  \centering
    \includegraphics[width=0.70\linewidth, trim= 0pt 0pt 0pt 22pt, clip ]{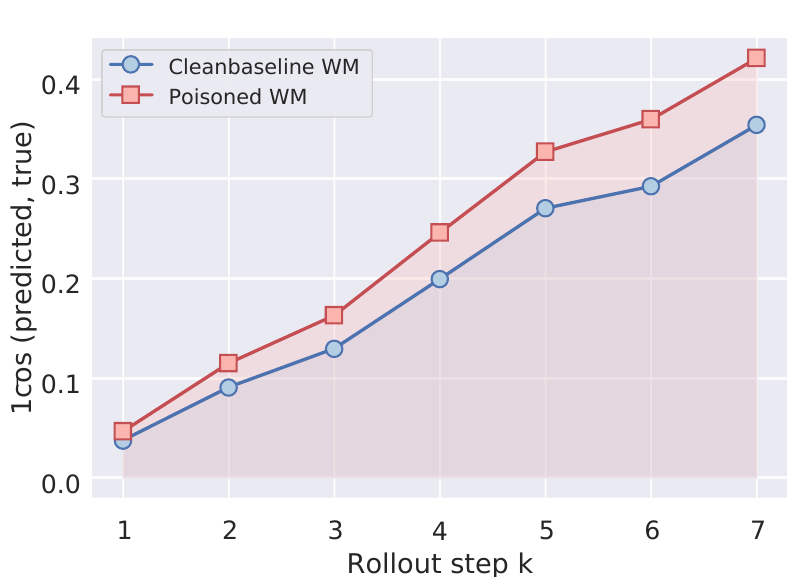}
  \caption{\textbf{Cumulative cosine drift over long-horizon imagination on clean data.} Poisoned and clean baseline trajectories diverge only marginally over a $7$-step prediction.}
  \label{fig:rollout_drift}
\end{figure}

\begin{figure}[t]
    \centering
    \includegraphics[width=0.95\linewidth, trim= 0pt 10pt 0pt 7pt, clip]{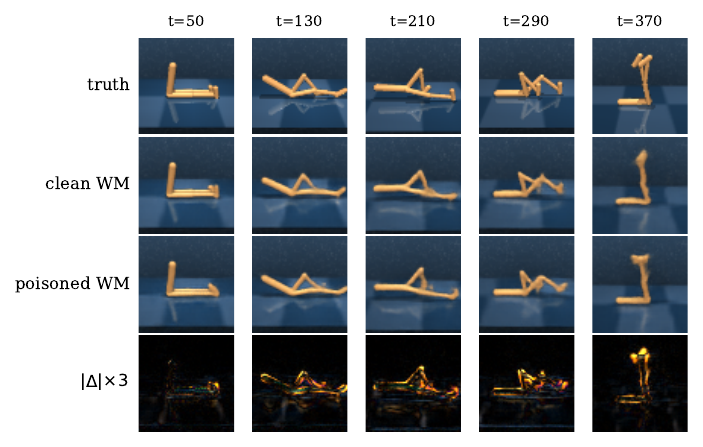}
    \caption{Stealth audit via predicted state reconstruction on Walker walk. Ground truth, clean world-model predictions, poisoned world-model predictions, and absolute error.}
    \label{fig:stealth_recon}
\end{figure}

A victim downloading a pretrained world model will typically begin by evaluating its reconstruction quality on clean observations, the plausibility of its short-horizon predictions, and the usability of downstream control on clean data. These are precisely the diagnostics considered here. Table~\ref{tab:stealth} evaluates these diagnostics across independently trained checkpoints, while Figures~\ref{fig:rollout_drift}-\ref{fig:stealth_recon} provide a paired view of prediction drift and reconstruction quality.

\begin{table}[t]
\centering
\caption{\textbf{Defender-side stealth audit across independently trained checkpoints.} Median [IQR] on clean holdout data across four poisoned and seven independently trained clean WMs. Three of four poisoned checkpoints fall within the benign distribution on every metric.}
\label{tab:stealth}
\small
\begin{tabular}{lcc}
\toprule
Metric & Poisoned ($n=4$) & Clean ($n=7$) \\
\midrule
Reconstruction \\
\quad MSE${\times}10^{-3}$
  & $1.026\,[0.896,\,1.789]$
  & $0.865\,[0.764,\,0.945]$ \\
\quad PSNR
  & $29.90\,[28.30,\,30.56]$
  & $30.63\,[30.26,\,31.18]$ \\
1-step\\
\quad MSE${\times}10^{-3}$
  & $2.258\,[1.963,\,3.644]$
  & $2.010\,[1.781,\,2.166]$ \\
\quad PSNR
  & $26.47\,[25.01,\,27.14]$
  & $26.97\,[26.65,\,27.50]$ \\
Pst-Pr KL
  & $19.36\,[17.31,\,25.15]$
  & $17.24\,[15.76,\,18.28]$ \\
\bottomrule
\end{tabular}
\end{table}

Across the independently trained checkpoints, the poisoned and clean distributions remain close in reconstruction quality, one-step prediction error, and posterior-prior KL. Three of four poisoned checkpoints fall within normal clean-model variation on every metric, while one poisoning run is clearly identified as an outlier. Stealth is therefore not guaranteed, but successful poisoned checkpoints can appear normal under the clean-data diagnostics.

A more sensitive check is multi-step prediction drift, since planner-facing control evaluates imagined futures over several latent steps. Figure~\ref{fig:rollout_drift} shows that the poisoned model does drift slightly more than the clean baseline, but the difference is small and does not separate the poisoned checkpoint from normal reseeding variability unless a paired clean reference is available. The same conclusion is visible in the reconstruction strip in Figure~\ref{fig:stealth_recon}: the poisoned predictions are somewhat noisier than the clean baseline, but they remain close enough to the ground truth that a victim would still consider the model deployable.

Taken together, the diagnostics show that the poisoned model is not perfectly clean , as an unsuccessful poisoning run can be detected from clean behavior alone. However, successful poisoned checkpoints can fall within normal clean-model variation on the same diagnostics a victim would run before deployment. Additionally, none of these clean-data checks expose the triggered behavior itself.


\section{Defenses}
\label{sec:defenses}
The following section analyzes whether standard post-hoc defenses can actually neutralize the backdoor behavior in the attacked world model. We consider two defender postures:  \emph{repair-time} defenses, which modify the suspect model on clean data before deployment or actor training; and \emph{deployment-time} defenses, which leave the model unchanged and instead focus on filtering suspicious inputs at inference. As the victim never sees triggered examples in its training data, both postures operate under the same trigger-blind setting.
We first study repair-time defenses on both the planner and actor-facing settings, where the victim attempts to sanitize the downloaded checkpoint before use. We then turn to deployment-time detection, where the goal is not to repair the model but to catch trigger-bearing inputs. Throughout, the DoS column reports the clean$-$triggered gap in each setting's task metric: success-rate percentage points (pp) for the planner and raw episode return degradation for the actor.

\subsection{Planner-facing Repair-time Defenses}
\label{subsec:repair-defenses}
We evaluate four defenses compatible with the trigger-blind threat model. \textbf{D1.~Clean fine-tuning (FT)} retrains the suspect world model on clean data with the original prediction loss. \textbf{D2.~Fine-pruning~\cite{liu2018finepruning}} ranks predictor-MLP hidden units by mean activation on clean validation data, zeroes the lowest $k$, and then fine-tunes the model. \textbf{D3.~Adversarial Neuron Pruning (ANP)~\cite{wu2021anp}} learns bounded perturbations on hidden units that maximize clean loss, prunes the top-$k$ units ranked by perturbation magnitude. \textbf{D4.~Neural Attention Distillation (NAD)~\cite{li2021nad}} first builds a teacher by briefly fine-tuning the suspect model on clean data, then retrains the original model with clean prediction loss plus an $\ell_2$ attention distillation term against the teacher. We exclude defenses that require labeled triggered data or a paired clean reference model, as both assumptions violate the threat model in Section~\ref{sec:threat-model}.

\begin{figure}
    \centering
    \includegraphics[width=0.95\linewidth, trim= 20pt 36pt 15pt 7pt, clip]{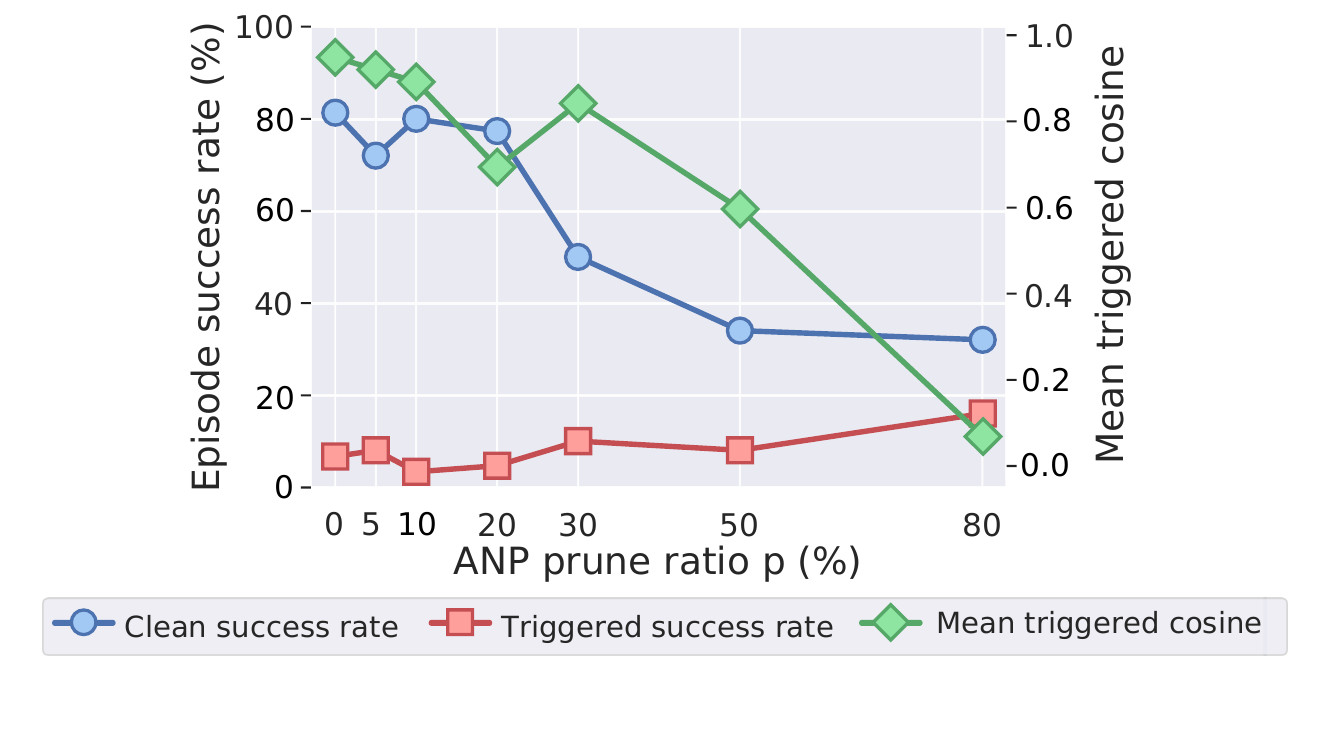}
    \caption{Effect of ANP on the planner-facing patch trigger across pruning ratios $p$.}
    \label{fig:anp_defense}
\end{figure}

Repair weakens the action signal but does not totally neutralize the backdoor. Table~\ref{tab:defenses-headline} reports the four defenses on the planner-facing attack. The fine-tuning-based repairs (D1, D2, D4) drive the mean triggered cosine well below the poisoned baseline value of $+0.948$, so on that single metric, they look effective. Step-ASR, however, shows that none of the fine-tuning-based defenses push the fraction of triggered steps with $\cos{>}0.8$ below $27\%$. Meaning that even after the defenses are applied, more than one-fourth of the taken actions in the presence of the trigger will be highly aligned with the target backdoor action. At low pruning ratios, ANP is unable to significantly decrease the backdoor's effectiveness, and only at aggressive ratios ($p{\geq}80\%$) does Step-ASR fall meaningfully, but with catastrophic loss of clean accuracy (see Figure~\ref{fig:anp_defense}). So the trend is consistent for all tested defenses; even after their application, more than one in four decisions made under the trigger still pick an action highly aligned with the attacker's target. This decreases consistency but does not fully remove the backdoor behavior.

The trigger still functions as a denial-of-service. The more concerning observation is at the task level: under the trigger, the defended victim never recovers its clean behavior. Across every repair in Table~\ref{tab:defenses-headline}, triggered task success stays at or below $10\%$, while the clean success rate of the same defended model remains $67$-$82\%$ (excluding the collapsed ANP $p{=}50$--$100\%$). The DoS gap (clean$-$triggered task success) is $\geq\!61$pp in every case. Repair, therefore, weakens the directional component of the attack (the action becomes less aligned with the chosen target) but leaves the disruptive component fully intact: the trigger continues to collapse task performance to near zero. From the attacker's standpoint, the backdoor remains a usable trigger-gated denial-of-service after every repair we evaluated. Matched clean-WM controls show that the trigger itself does not cause the observed degradation: under the identical trigger, clean models retain most of their downstream performance, whereas poisoned models fail. Therefore, the residual triggered failure after repair is attributable to the backdoor rather than to ordinary OOD sensitivity (Appendix~\ref{app:clean_controls}).

\begin{table}[t]
\centering
\caption{Repair-time defenses on the planner-facing setting (Reacher). \textbf{C.SR}: clean success rate. \textbf{T.SR}: triggered success rate. \textbf{T.Cos}: mean action-target cosine on triggered inputs. \textbf{Step-ASR}: fraction of triggered steps with $\bar c{>}0.8$. \textbf{DoS}: clean$-$triggered task success (pp).}
\label{tab:defenses-headline}
\small
\setlength{\tabcolsep}{4pt}
\begin{adjustbox}{max width=\columnwidth}
\begin{tabular}{@{}l ccccc@{}}
\toprule
\textbf{Defense} & \textbf{C.SR} & \textbf{T.SR} & \textbf{T.Cos} & \textbf{Step-ASR} & \textbf{DoS}\\
\midrule
No defense                & $78\%$ & $10.6\%$ & $+0.948$ & $95.0\%$ & $67$pp \\
\midrule
D1.\ Clean FT             & $81\%$ & $\phantom{0}8\%$ & $+0.399$ & $39.2\%$ & $73$pp \\
D2.\ Fine-prune~$10\%$ & $82\%$ & $\phantom{0}7\%$ & $+0.388$ & $38.7\%$ & $75$pp \\
D3.\ ANP $p{=}10\%$ & $80\%$ & $\phantom{0}3\%$ & $+0.891$ & $86.3\%$ & $77$pp \\
\hspace{1.8em}ANP $p{=}20\%$ & $77\%$ & $\phantom{0}5\%$ & $+0.694$ & $61.5\%$ & $73$pp \\
\hspace{1.8em}ANP $p{=}80\%$ & $32\%$ & $16\%$ & $+0.067$ & $23.1\%$ & $16$pp \\
D4.\ NAD                  & $67\%$ & $\phantom{0}5\%$ & $+0.220$ & $27.2\%$ & $61$pp \\
\bottomrule
\end{tabular}
\end{adjustbox}
\end{table}


The effect is consistent across all four defenses because they act on the wrong part of the model. Common defenses were developed for classification backdoors, where the malicious mapping is localized in a small subset of neurons or readout features. The world-model backdoor studied here is different: the trigger routes the observation into a malicious latent trajectory, and the downstream controller then amplifies that trajectory through its own optimization. Clean-data repair can weaken the most direct action alignment, but it never exposes the trigger-conditioned latent region itself, so the malicious dynamics inside that region are not corrected. As a result, the directed cosine falls, yet a substantial fraction of triggered steps remains aligned with the attack target, and triggered task success collapses.

\subsection{Actor-facing Repair-time Defenses}
\label{subsec:repair-defenses-actor}

We replicate the same protocol on the actor-facing Dreamer (Walker) setting, where the suspect artifact is the poisoned world model and the victim trains an actor on top of it after applying the defense. The threat model and trigger are identical to the planner-facing case. Each row of Table~\ref{tab:defenses-walker} is evaluated by training an actor for $160{,}000$ environment-equivalent steps on the defended world model and reporting clean and triggered episodic returns together with the Step-ASR at the same $0.8$ cosine threshold used in the planner-facing table.


\begin{table}[t]
\centering
\caption{Repair-time defenses on the actor-facing setting (Walker). \textbf{C.Ret}: clean episode return. \textbf{T.Ret}: triggered episode return. \textbf{T.Cos}: mean action-target cosine on triggered inputs. \textbf{Step-ASR}: fraction of triggered steps with T.Cos$>0.8$. \textbf{DoS}: clean$-$triggered task success.}
\label{tab:defenses-walker}
\small
\setlength{\tabcolsep}{4pt}
\begin{adjustbox}{max width=\columnwidth}
\begin{tabular}{@{}l ccccc@{}}
\toprule
\textbf{Defense} &\textbf{C.Ret} & \textbf{T.Ret} & \textbf{T.Cos} & \textbf{Step-ASR} & \textbf{DoS}\\
\midrule
No defense                & $961$ & $\phantom{0}42$ & $+1.00$ & $100.0\%$ & $919$\\
\midrule
D1.\ Clean FT             & $946$ & $\phantom{0}24$ & $+1.00$ & $100.0\%$   & $922$ \\
D2.\ Fine-prune~$10\%$ & $947$ & $\phantom{0}42$ & $+1.00$ & $100.0\%$      & $905$ \\
D3.\ ANP $p{=}10\%$       & $564$ & $\phantom{0}10$ & $+0.69$ & $\phantom{00}0.0\%$   & $554$ \\
\hspace{1.8em}ANP $p{=}20\%$ & $920$ & $\phantom{0}22$ & $+0.12$ & $\phantom{00}0.0\%$  & $898$ \\
\hspace{1.8em}ANP $p{=}80\%$ & $197$ & $\phantom{00}6$ & $+0.16$ & $\phantom{00}0.0\%$  & $191$ \\
D4.\ NAD                  & $953$ & $\phantom{0}42$ & $+1.00$ & $100.0\%$   & $911$ \\
\bottomrule
\end{tabular}
\end{adjustbox}
\end{table}

The actor-facing picture is substantially different from the planner-facing one in Table~\ref{tab:defenses-headline}. Every fine-tuning-focused defense (D1, D2, D4) recovers a clean return of $\geq\!946$ but does not affect the targeted attack. We also analyze what happens when the victim performs a stronger clean fine-tuning across learning rates and clean-data budgets (see Appendix~\ref{app:clean_ft}). At a learning rate of $10^{-4}$, the repaired poisoned model still returns only $17$ per triggered episode versus $952$ on clean inputs, while an identically repaired clean world model returns $955$ and $957$, respectively. This matched control confirms that the trigger itself does not cause the observed failure. Increasing the repair strength can eventually remove the triggered behavior, but only in the aggressive setting: at $3\times\!10^{-3}$, the poisoned model reaches comparable triggered and clean returns, but at the expense of the latter one ($648$ vs. $670$). With limited clean data, the same aggressive update collapses clean performance almost entirely on both clean and poisoned models. Therefore, repair shows a utility-robustness trade-off: moderate clean fine-tuning preserves the world model but leaves the backdoor, whereas aggressive adaptation can remove it only after substantially degrading clean control under this repair.

\subsection{Deployment-time Detection}
\label{subsec:deployment-detection}

Another defense angle is to intercept suspicious inputs before giving a response. We evaluate STRIP~\cite{gao2019strip}: an incoming observation is blended with a set of clean reference frames, and the consistency of the model's response across those blends is used as a detection score. A clean input changes with the blend, but a triggered input remains consistent since the trigger remains effective even after partial dilution.

\begin{figure*}[t]
    \centering
    \begin{subfigure}[t]{0.32\textwidth}
        \centering
        \includegraphics[width=\linewidth, trim= 0pt 8pt 0pt 7pt, clip]{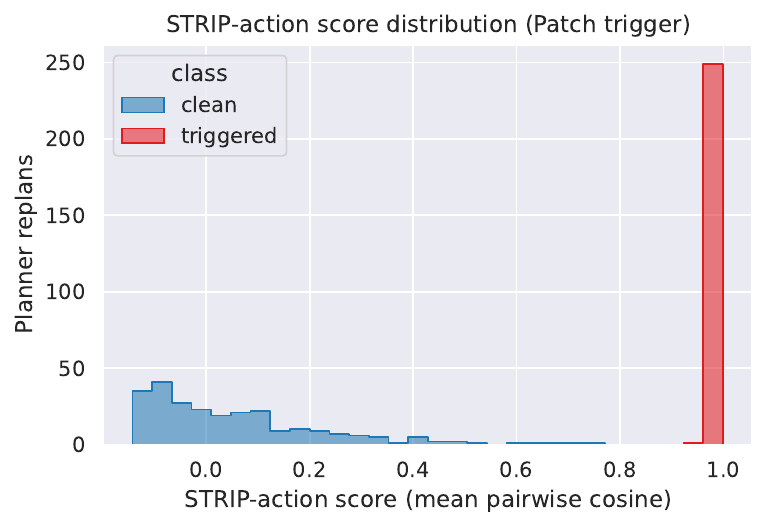}
        \caption{STRIP-action, patch trigger. Triggered inputs stay consistent across blends and score near one, while clean inputs scatter lower; $\mathrm{AUC}=1.00$.}
        \label{fig:strip_action_patch}
    \end{subfigure}\hfill
    \begin{subfigure}[t]{0.32\textwidth}
        \centering
        \includegraphics[width=\linewidth, trim= 0pt 8pt 0pt 7pt, clip]{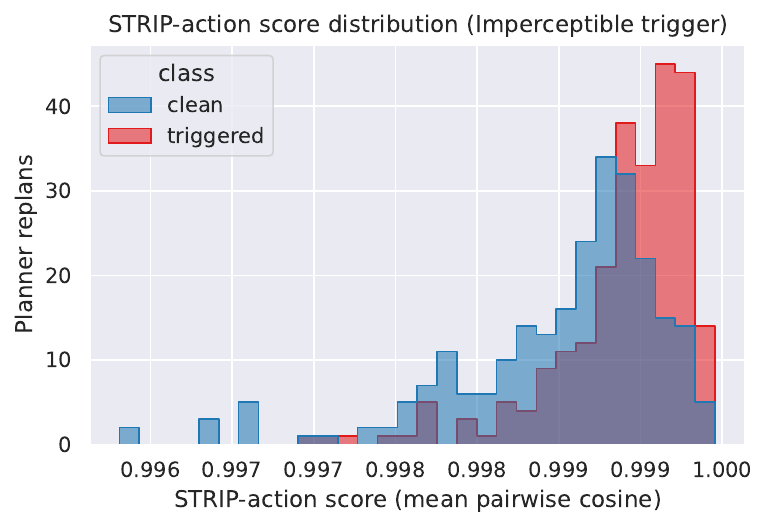}
        \caption{STRIP-action, additive trigger. The trigger is washed out at the planner head, so the distributions overlap and detection collapses to $\mathrm{AUC}=0.72$.}
        \label{fig:strip_action_imperceptible}
    \end{subfigure}\hfill
    \begin{subfigure}[t]{0.32\textwidth}
        \centering
        \includegraphics[width=\linewidth, trim= 0pt 8pt 0pt 7pt, clip]{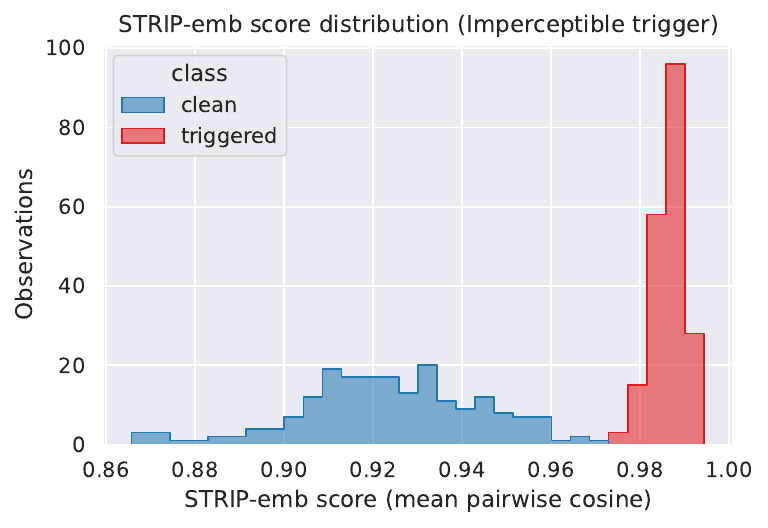}
        \caption{STRIP-emb, additive trigger. Applying the same consistency test at the encoder output recovers clean separation; $\mathrm{AUC}=1.00$.}
        \label{fig:strip_encoder_imperceptible}     
    \end{subfigure}
    \caption{\textbf{STRIP detection: action-level vs.\ encoder-level scoring.} Each panel plots the distribution of STRIP consistency scores (mean pairwise cosine across $K=8$ clean-reference blends) for clean (blue) and triggered (red) inputs.}
    \label{fig:strip_hist}
\end{figure*}
Applied at the planner's first action, STRIP-action scores each observation by the mean pairwise cosine of the first actions produced across $K{=}8$ blends $\tilde x_i = \alpha x + (1-\alpha)b_i$, $\alpha{=}0.5$, with partners drawn from a held-out clean pool:
\begin{equation*}
s_{\mathrm{strip\text{-}act}}(x) = \tfrac{1}{K(K-1)} \sum_{i \neq j} \cos\!\big(a_1(\tilde x_i),\, a_1(\tilde x_j)\big).
\end{equation*}
On the patch trigger, triggered scores concentrate near one while clean scores scatter below, giving AUC\,$=1.00$ (Figure~\ref{fig:strip_action_patch}). But the detection efficacy is not uniform across trigger families. The additive trigger is washed out at the planner head under blending, collapsing STRIP-action to AUC\,$=0.72$ (Figure~\ref{fig:strip_action_imperceptible}).

However, another concern comes from the detection cost. Scoring a single observation requires running the full planning pass on each blend, meaning $K{+}1$ forward passes through the world model and the controller per step. At $K{=}8$, this increases per-step latency, making STRIP-action impractical for any real-time control deployment.

Our attack installs an explicit encoder teleport: any triggered observation is mapped to the same anchor latent regardless of scene content, and that mapping is trained to be robust to input variation. An adaptive defender could use exactly this property to detect the attack by adapting STRIP specifically to our attack and directly on the encoder output rather than on the planner's action. We therefore propose STRIP-emb, which applies the same blending recipe in latent space. Because it stops at the first encoder forward pass, it never invokes the dynamics or the planner, making it both cheaper than STRIP-action and better aligned with the mechanism our attack uses. On the patch trigger, STRIP-emb matches STRIP-action at AUC\,$=1.00$ at a fraction of the cost. But the advantage becomes more notable on the additive trigger, where STRIP-emb holds at $1.00$ (Figure~\ref{fig:strip_encoder_imperceptible}).  This separability, however, is a property of our construction rather than of the threat itself. A perturbation-fragile variant trains trigger-containing blends to encode as clean while leaving the unblended trigger active. This reduces STRIP-emb from AUC$=1.00$ to $0.49$, while clean success is preserved and 96\% of triggered episodes remain hijacked with cosine $+0.87$.

We additionally evaluate an inversion-based detector by adapting DECREE~\cite{Feng_2023_CVPR}. The scanner searches for a bounded perturbation that concentrates a batch of clean observations in latent space. When the defender is given the attacker's anchor $z^\star$, the recovered perturbation separates poisoned from clean world models: the anchor cosine ranges from $ 0.51$ to $0.89$ on poisoned models and from $ 0.03$ to $0.20$ on clean ones. However, $z^\star$ would not be available to the victim, and in this case, the detector becomes substantially weaker. Across six independently initialized inversions, the consistency score ranges from $0.66$-$0.95$ for poisoned models and $0.28$-$0.95$ for benign ones. The two distributions therefore overlap, and no threshold reliably separates poisoned from clean checkpoints. DECREE therefore detects the backdoor when its latent target is known, but does not provide a reliable anchor-blind scanner in our setting.


A reliable detector still leaves the harder problem open. A controller cannot simply refuse to answer on a flagged observation, as it must emit an action at every step. Prediction-observation consistency is another promising world-model-specific signal. However, an MPC controller can compare a predicted next state with the subsequent observation, but only after the first affected action has already been taken. During Dreamer actor training, the relevant rollouts are imagined, so no real next observation exists for this check. Therefore, in both cases, a safe fallback is still needed until new observations arrive.




\section{Related Work}
\label{sec:related-work}
Backdoor attacks were introduced in the context of image classification by Gu et al.~\cite{gu2019badnets}, who showed that a small patch trigger inserted into a fraction of training samples and paired with a malicious label causes a deployed classifier to misclassify any triggered input. Chen et al.~\cite{chen2017targeted} extended this to blended triggers, and a subsequent line of work generalized the attack to invisible triggers~\cite{nguyen2021wanet}, semantic triggers~\cite{li2021invisible}, and label-consistent variants~\cite{turner2019labelconsistent}. In all cases, the malicious behavior is encoded directly in the classifier's input–label mapping: the training loop is presented with corrupted (input, label) pairs and the classifier learns them. Backdoors against reinforcement-learning policies have been studied via reward poisoning~\cite{kiourti2020trojdrl,yang2019design,wang2021backdoorl}, observation poisoning~\cite{wang2021stop,gong2022mind}, and communication backdoors in multi-agent settings~\cite{chen2022marnet}. These assume the attacker can modify the victim's training loop by poisoning the environment, reward, or trajectories. Test-time adversarial perturbations on RL policies~\cite{huang2017adversarial,lin2017tactics,gleave2020adversarial} degrade performance without training-time compromise and are typically untargeted. Our threat model removes training-loop access entirely: the attacker compromises only a pretrained world model, the victim trains a fresh controller on clean data, and the trigger is a fixed observation transform at deployment.

Beyond classification backdoors, recent work has examined supply-chain attacks against pretrained models: backdoors injected during language-model pretraining~\cite{wallace2020concealed}, attacks on contrastive vision encoders~\cite{carlini2022poisoning,jia2022badencoder}, and risks from compromised models on public hubs~\cite{zhang2025transtroj}. INFUSE~\cite{zhou2026infuse} shows that backdoors injected into vision–language–action base models can survive extensive downstream fine-tuning by targeting fine-tune-insensitive modules. We differ from INFUSE in both the artifact and the mechanism: the compromised backbone is the latent dynamics rather than a vision–language–action policy, and the malicious behavior is not stored as a fixed input–output mapping that must survive fine-tuning, but reconstructed from the poisoned dynamics by the victim's own controller optimization. The shared conclusion is that, as practitioners increasingly download large pretrained models rather than train from scratch, the model supply chain becomes a primary attack surface. Latent world models for control are at an earlier point in this trajectory than language, vision, or VLA backbones, but the same pattern is emerging: the artifacts are expensive to train, are shared across teams, and serve as reusable backbones. Our threat model instantiates a supply-chain attack in the latent-world-model regime, where malicious behavior must survive downstream controller training on clean data.

Recent work has begun to attack learned dynamics and sequential decision systems under related threat models. BadEncoder~\cite{jia2022badencoder} establishes trigger-conditioned representation routing in pretrained encoders, which is closely related to our encoder teleport. Daze~\cite{rathbun2026bewareuntrustedsimulators} studies reward-independent dynamics manipulation in an untrusted simulator, while SleeperNets~\cite{rathbun2024sleepernets} and TrojanTO~\cite{DBLP:journals/corr/abs-2506-12815} study targeted action or trajectory backdoors in reinforcement-learning systems. Our actor-facing construction builds on these primitives rather than introducing representation routing or reward-independent dynamics manipulation themselves. SWAAP~\cite{hu2026stealthy} also manipulates a learned world model, but assumes access to the world-model training process and poisons online transitions to steer an updating model toward a task-specific worst-case target that degrades planning performance. Parmar~\cite{parmar2026safetyrisks} demonstrates trajectory-persistent attacks on GRU-based RSSMs and a DreamerV3 checkpoint under adversarial fine-tuning, again with attacker access to later training. In a different domain, backdoors in continuous latent reasoning show that perturbing a single embedding can hijack long-horizon latent trajectories while evading token-level defenses~\cite{parekh2026thoughtsteer}. In contrast, our attacker releases a poisoned world-model checkpoint and loses access before the victim trains a fresh controller on completely clean data. The planner setting adds a separate constraint, as representation routing or actor-style dynamics shaping alone does not determine the output of CEM. Removing $\mathcal{L}_{\text{cem-plan}}$ drops the triggered cosine from $+0.95$ to $+0.14$ (Table~\ref{tab:ablation-main}).

\section{Conclusion}
\label{sec:conclusion}

We present, to our knowledge, the first checkpoint-only backdoor that installs targeted control behavior into a latent world model and survives clean downstream controller training: it hijacks the controller by poisoning only the released world model, while the victim's data, controller, objective, and evaluation remain clean. In this supply-chain setting, the attacker controls only the world-model checkpoint, while the victim controls the controller, the task objective, the data, and the evaluation procedure. The malicious behavior is therefore never written explicitly as a trigger-to-action rule. It must be encoded in the learned dynamics and later recovered by the downstream controller through its own optimization, whether via policy-gradient training in imagination or cost-ranked planning at deployment.

Across both the planner-facing and actor-facing settings we study, the attack steers downstream behavior toward the attacker's chosen action while preserving clean utility on untriggered inputs, and it does so for trigger families ranging from a conspicuous patch to a subtle global perturbation or a realistic object/actor color change. Matched clean-WM controls show that the same trigger does not reproduce the targeted behavior in benign checkpoints, confirming that the hijack is induced by the poisoned world model rather than by the trigger alone. The effect is also trigger-gated: the hijack appears when the trigger is present and largely reverts to clean behavior once it is removed. Across the diagnostics we evaluate, three of four poisoned checkpoints fall within normal benign variation, showing that a compromised checkpoint can remain difficult to distinguish from independently trained clean models before deployment.

The backdoor is also durable under a range of evaluated repairs, although its persistence depends on the victim's adaptation budget. Moderate clean-data fine-tuning preserves clean utility but can leave the triggered failure intact, whereas sufficiently aggressive adaptation can remove the malicious behavior at the cost of degrading clean control in the end-to-end setting, particularly when clean data is limited. Deployment-time detection is similarly incomplete. STRIP detects our standard construction efficiently at the encoder, but an adaptive perturbation-fragile variant reduces its AUC to chance while preserving the attack. An adapted DECREE scanner separates poisoned from clean models when the malicious anchor is known, but becomes substantially weaker when that anchor is unavailable. However, even a reliable detector does not restore control: a controller must still take an action at every step.

These results identify the world-model backbone as a security-critical artifact on its own: a victim can keep the data, controller, and evaluation protocol entirely clean and still inherit a backdoor through the supplied checkpoint. We demonstrate this on DreamerV3 and LeWorldModel in standard simulated control tasks, the setting in which the supply-chain mechanism can be isolated and studied directly. The vulnerability we study arises from the latent interface these models share: an observation is encoded into a latent state whose predicted dynamics the controller then trusts, and the condition that makes the attack possible is the reuse of a pretrained world-model checkpoint rather than training one from scratch. This becomes more relevant precisely as world models grow more expensive to train and reuse becomes more attractive. We therefore expect this concern to become increasingly important as the field moves toward larger video and joint-embedding world models, and we argue that shared world-model checkpoints should be treated like other security-sensitive supply-chain components rather than as neutral dynamics backbones.
\bibliographystyle{IEEEtran}
\bibliography{references}
%


\appendix

\section{Loss-Component Ablations}
\label{app:ablations}

We ablate each loss term in isolation, keeping all others at their default weights, and re-run the full downstream pipeline. Table~\ref{tab:ablation-main} summarizes the effect of removing each term; below we give the per-term reading and the encoder-routing rows specific to the planner (Table~\ref{tab:abl-lewm}).

\subsection{Planner-Facing Attack Component Ablation}
\label{app:abl-lewm}


\begin{table}[ht]
\centering
\caption{\textbf{Encoder routing on Reacher} (planner). Rows beyond the main-text ablation (Table~\ref{tab:ablation-main}): removing $\mathcal{L}_{\text{enc}}$ under a visible patch vs.\ a subtle additive trigger, and removing the dynamics-side route to $z^\star$ as well (\emph{no\_anchor}). The full attack is repeated for reference.}
\label{tab:abl-lewm}
\small
\setlength{\tabcolsep}{5pt}
\renewcommand{\arraystretch}{1.1}
\begin{adjustbox}{max width=\columnwidth}
\begin{tabular}{@{}lcccc@{}}
\toprule
Ablation & Clean SR & Trig SR & Trig Cos & Step-ASR \\
\midrule
Full attack             & $78\%$ & $10.6\%$ & $+0.95$ & $95\%$   \\
no\_enc (patch) & $82\%$ & $0\%$    & $+0.82$ & $64.9\%$ \\
no\_enc (additive)      & $76\%$ & $2\%$    & $-0.21$ & $12\%$   \\
no\_anchor              & $76\%$ & $8\%$    & $-0.34$ & $6\%$    \\
\bottomrule
\end{tabular}
\end{adjustbox}
\end{table}


The encoder teleport behaves differently in the two settings, and the planner case is the subtle one. For a visible trigger, \emph{no\_enc} leaves the attack largely functional ($64.9\%$ Step-ASR, cosine $+0.82$): since the world model remains present at deployment, the loop and CEM-plan terms can shape the dynamics, allowing the predictor to route the visible patch toward $z^\star$ on its own. This fails once the trigger becomes subtle. With the additive trigger, \emph{no\_enc} drops to $12\%$ Step-ASR and the cosine flips to $-0.21$, showing that the dynamics cannot reliably route a subtle trigger without the explicit encoder-side teleport. The \emph{no\_anchor} row removes both routes to $z^\star$ and reduces Step-ASR further to $6\%$. The encoder teleport is therefore what makes the attack \emph{trigger-agnostic}: a visible trigger can sometimes be recovered directly by the poisoned dynamics, whereas a subtle trigger requires the encoder to provide the bridge to the same fixed $z^\star$.

\subsection{Actor-Facing Attack Component Ablation}
\label{app:abl-dreamer}

We poison the Dreamer world model on Walker-walk and train a fresh actor in imagination for 250{,}000 environment-equivalent steps (Table~\ref{tab:ablation-main}). The actor reading is sharper than on Reacher. \emph{no\_enc} neutralizes the attack outright (cosine $-0.31$, Step-ASR $\approx 0\%$, triggered return equal to clean): at deployment, the world model is discarded, so with no teleport the trigger image encodes to a latent the actor does not recognize, cutting the only bridge to $z^\star$. \emph{no\_loop} still destabilizes the walker (triggered return $41$) but the action is incoherent ($\cos = +0.26$, Step-ASR $0\%$): the teleport reaches $z^\star$ but the dynamics no longer hold imagination there under $a^\star$, so the attack is disruptive without being targeted. \emph{no\_trap} collapses the attack ($\cos \approx 0$) and drops clean utility to $743$, since without a low-value sink the actor has no return-difference signal singling out $a^\star$. \emph{no\_sticky} still fires ($\cos = +0.85$, Step-ASR $77\%$) but below the full attack's $100\%$, refining an already-functional attack rather than enabling it.

Across both settings, the encoder teleport is the binding link to $z^\star$, but the effect of removing it differs: on the actor it is the only bridge, so the attack dies; on the planner the dynamics substitute for a visible trigger, and only when the trigger is imperceptible, or the anchor itself is removed, does the planner collapse to the actor's picture. The remaining terms shape behavior at $z^\star$, the loop installs the fixed point, the trap the value contrast, the sticky term the inescapability, and $\mathcal{L}_{\text{cem-plan}}$ (Appendix~\ref{app:softcem}) aligns the dynamics with the deployed planner.

\section{The softCEM Planner Surrogate and Stability Regularizers}
\label{app:planner-stability}

The planner attack uses a differentiable surrogate of the deployment CEM planner (Appendix~\ref{app:softcem}) and two training-procedure regularizers (Appendix~\ref{app:stop-gradient}--\ref{app:teacher}) that keep the attack from leaking into clean rollouts on harder environments:
$
\mathcal{L}_{\text{stab}}(\theta)
= \mathcal{L}_{\text{sg}}(\theta) + \lambda_{\text{tea}}\,\mathcal{L}_{\text{teacher}}(\theta).$
Neither stability term is a conceptual component of the attack: the encoder teleport, stabilizing loop, and CEM-aware plan loss carry the malicious behavior. The role of the regularizers is to keep that malicious behavior from corrupting the model's behavior on clean inputs during training.

\subsection{The softCEM Surrogate}
\label{app:softcem}

The deployment-time controller in LeWorldModel is a sampling-based MPC: at every decision step, the planner maintains a Gaussian 
over action sequences 
, samples $N$ candidates, rolls each candidate forward through the world model, ranks them by a goal cost 
, refits the Gaussian to the top-$K$ elites, repeats for $S$ optimization steps, and executes the first action of the final elite mean. CEM is non-differentiable (top-$K$ selection is a hard threshold), and a naive single-step surrogate optimized only against the first sampled candidate does not match what the deployed planner ends up executing, which is the \emph{mean} of an elite distribution after several refits.

To train the world-model parameters $\theta$ against this end-to-end behavior, we use a soft, differentiable variant we refer to as \emph{softCEM}. It preserves the iterative structure of CEM but (i) replaces the hard top-$K$ elite selection with a softmax weighting and (ii) detaches all but the last iteration so that gradients flow only through the final refit, keeping memory bounded over $S$ steps.

\paragraph{Iterative refits} Initialize $\mu_0 = \mathbf{0}_{H \times A}$ and $\sigma_0 = \sigma_{\text{init}} \mathbf{1}_{H \times A}$ (we use $\sigma_{\text{init}}=1$). For each iteration $s = 1, \dots, S$ and each (triggered context, goal) pair indexed by $p$:
\begin{align}
A^{(p,n)}_s &\sim \mathcal{N}(\mu^{(p)}_{s-1},\, (\sigma^{(p)}_{s-1})^2 \mathbf{I}),
  \quad n = 1,\dots,N, \\
A^{(p,0)}_s &= \mu^{(p)}_{s-1}, \\
c^{(p,n)}_s &= \big\lVert \hat z_H\!\big(A^{(p,n)}_s;\, T(h^{(p)})\big) - z_{g^{(p)}} \big\rVert_2^2, \\
w^{(p,n)}_s &= \mathrm{softmax}_n\!\Big(\!-\,\tfrac{c^{(p,n)}_s - \min_{n'} c^{(p,n')}_s}{\tau}\Big), \label{eq:softcem-weights}\\
\mu^{(p)}_s &= \sum_n w^{(p,n)}_s\, A^{(p,n)}_s, \\
(\sigma^{(p)}_s)^2 &= \sum_n w^{(p,n)}_s\, \big(A^{(p,n)}_s - \mu^{(p)}_s\big)^2 + \sigma^2_{\min},
\end{align}
where $T(h^{(p)})$ is the triggered context (real prefix + triggered last frame), $z_{g^{(p)}}$ is the goal embedding, $\tau$ is the softmax temperature, and $\sigma^2_{\min}$ is a floor that prevents collapse. For $s < S$, we detach $\mu^{(p)}_s$ and $\sigma^{(p)}_s$ before the next iteration: only the candidates sampled at the final step $s = S$ produce gradients into $\theta$.

Writing $A^\star = (a^\star, \dots, a^\star)$ for the target plan that repeats the target action $a^\star$ over the horizon and denoting the final-iteration weights and mean simply as $w^{(p,n)}$ and $\mu^{(p)}$, we combine three terms:
\begin{align}
\mathcal{L}^{(p)}_{\text{exp}} &= \sum_n w^{(p,n)}\,\tfrac{1}{HA}\big\lVert A^{(p,n)}_S - A^\star \big\rVert_2^2, \\
\mathcal{L}^{(p)}_{\text{mean}} &= \tfrac{1}{HA}\big\lVert \mu^{(p)} - A^\star \big\rVert_2^2, \\
\mathcal{L}^{(p)}_{\text{cos}} &= 1 - \cos\!\big(\mathrm{vec}(\mu^{(p)}),\, \mathrm{vec}(A^\star)\big),
\end{align}
and average across $P$ context-goal pairs:
\begin{equation}
\mathcal{L}_{\text{cem-plan}}(\theta)
= \tfrac{1}{P}\sum_p \Big[ \lambda_{\text{exp}}\,\mathcal{L}^{(p)}_{\text{exp}}
                       + \lambda_{\text{mean}}\,\mathcal{L}^{(p)}_{\text{mean}}
                       + \lambda_{\text{cos}}\,\mathcal{L}^{(p)}_{\text{cos}} \Big].
\label{eq:cem-plan-loss}
\end{equation}
$\mathcal{L}_{\text{exp}}$ is the expected per-coordinate distance of an elite-weighted plan from $A^\star$ and provides the dominant gradient near high-cost candidates; $\mathcal{L}_{\text{mean}}$ and $\mathcal{L}_{\text{cos}}$ shape the refit elite mean itself, which is what the deployed planner actually executes (LeWM defaults to receding horizon equal to plan horizon, so the entire horizon is executed before replanning).

A surrogate that picks $\arg\min_n c^{(p,n)}_s$ at a single iteration trains only one path through one sampled candidate. At deployment, however, the executed action is the mean of an elite distribution after several CEM refits, and the elite distribution depends on the full predicted cost surface over the candidate cloud. Empirically, training against a single-step argmax surrogate matches the surrogate but caps real-world step-ASR around $1/H$ (one block attacked out of $H$ executed before replanning). Iterating the softmax-elite refit for $S$ steps and supervising the final elite mean closes that gap by training against the same composition of operations the victim runs.

\subsection{Action-Encoder Stop-Gradient}
\label{app:stop-gradient}

The world model contains a small sub-network, the \emph{action encoder}, that maps each candidate action into the latent space before the dynamics step. Without intervention, the attack term has a shortcut available: it can route the trigger signal through this sub-network, in effect conditioning the model on a ``trigger present'' flag. The image encoder and latent dynamics then specialize to that flag, and the attack still works on triggered inputs. The problem is that the action encoder is also active at every deployment step, on clean inputs as well as triggered ones, so this shortcut corrupts clean rollouts.

We block the shortcut by partitioning the parameters as $\theta = (\theta_e, \theta_a, \theta_d)$ into image encoder, action encoder, and latent dynamics, and applying a stop-gradient on $\theta_a$ in the backward pass of every attack term. The clean prediction loss continues to update $\theta_a$ as usual, so the action encoder is shaped only by clean data. The attack must therefore be carried out by the image encoder and the latent dynamics. In the planner objective, the attack term is $\mathcal{L}_{\text{atk}}\!\big(\theta_e,\,\text{sg}[\theta_a],\,\theta_d\big),$
where $\text{sg}[\cdot]$ is the stop-gradient operator.

\subsection{Multi-Step Frozen-Teacher Distillation}
\label{app:teacher}

The clean prediction loss in LeWorldModel is a one-step JEPA objective: predict the next-step latent given the history. This is too short-sighted to catch the drift the attack induces. The image encoder is being pulled toward $z^\star$ by the trigger-side terms, and in harder environments, that pull leaks into clean predictions over 4--7 latent steps, even though one-step predictions remain accurate. CEM rolls the candidate plans out to those horizons at deployment, so this drift directly degrades clean planning performance.

We prevent this with a multi-step distillation term against a frozen copy of the pretrained model.
\begin{equation}
\mathcal{L}_{\text{teacher}}(\theta)
= \frac{1}{H}\sum_{t=1}^{H}
\big\lVert z_t^{\text{stu}}(\theta) - \text{sg}\!\big[z_t^{\text{tea}}(\theta^\star)\big] \big\rVert_2^2.
\end{equation}
The rollout is autoregressive on each side: at each step, the predicted embedding is appended to the history that feeds the next prediction. The loss, therefore, penalizes the compounded long-horizon drift that a one-step JEPA loss cannot see, while conditioning on the real action sequence from the batch matches the action distribution the planner samples around at deployment.

The two regularizers are complementary. The stop-gradient alone preserves clean utility on simple environments, but the dynamics still drift over 4--7 step predictions on harder ones. Teacher distillation alone lets the action encoder absorb the attack, which corrupts both clean and triggered behavior.

\begin{table}[t]
\centering
\caption{\textbf{Effect of the stability regularizers} across the four LeWorldModel environments. }
\label{tab:stability-summary}
\small
\setlength{\tabcolsep}{4pt}
\begin{tabular}{ll rrr}
\toprule
Env. & Config & Clean SR\ & Trig.\ cos & Step-ASR \\
\midrule
\multirow{2}{*}{Reacher}
 & w/o stab. & $68\%$ & $+0.845$ & $80.0\%$ \\
 & w/ stab.  & $80\%$ & $+0.955$ & $98.6\%$ \\
\midrule
\multirow{2}{*}{TwoRoom}
 & w/o stab. & $36\%$ & $+0.674$ & $77.4\%$ \\
 & w/ stab.  & $84\%$ & $+0.899$ & $86.1\%$ \\
\midrule
\multirow{2}{*}{PushT}
 & w/o stab. & $62\%$ & $+0.893$ & $86.4\%$ \\
 & w/ stab.  & $82\%$ & $+0.891$ & $84.2\%$ \\
\midrule
\multirow{2}{*}{Cube}
 & w/o stab. & $62\%$ & $+0.732$ & $52.8\%$ \\
 & w/ stab.  & $70\%$ & $+0.727$ & $47.5\%$ \\
\bottomrule
\end{tabular}
\end{table}

\section{Controller-Transfer Robustness}
\label{app:controller-robustness}
\subsection{Planner Hyperparameters}
\label{app:planner-robustness}

The planner attack is trained against the differentiable softCEM surrogate of Appendix~\ref{app:softcem} ($N{=}256$, softmax selection, $S{=}5$), whereas the victim deploys the non-differentiable hard top-$K$ CEM planner of LeWorldModel and is free to choose its population size $N$, elite count $K$, and iteration budget $S$ (Section~\ref{sec:threat-model}). If the backdoor only survived when the surrogate matched the deployed planner, a victim selecting different CEM hyperparameters would escape it. We rule this out by poisoning a single Reacher checkpoint against the surrogate and evaluating it, unchanged, against the deployed planner across a $4\times$ range of $N$, a $6\times$ range of $K$, and a $10\times$ range of $S$ (Table~\ref{tab:planner-robustness}); every planner setting in the table is a choice the victim makes after release.

\begin{table}[t]
\centering
\caption{Planner-transfer robustness on Reacher. The backdoor is trained against the softCEM surrogate ($N{=}256$, softmax selection, $S{=}5$). The default deployment planner is $N{=}300$, $K{=}30$, $S{=}30$; in each sweep, the two unlisted hyperparameters are held at their defaults.}
\label{tab:planner-robustness}
\small
\setlength{\tabcolsep}{5pt}
\begin{adjustbox}{max width=\columnwidth}
\begin{tabular}{lcccc}
\toprule
Planner config & Clean SR & Trig SR & Trig cos & Step-ASR \\
\midrule
default & $78\%$ & $10.6\%$ & $0.948$ & $95\%$ \\
\midrule
$N{=}128$ & $74\%$ & $2\%$ & $0.970$ & $93\%$ \\
$N{=}512$ & $82\%$ & $0\%$ & $0.938$ & $83\%$ \\
$K{=}10$  & $68\%$ & $4\%$ & $0.954$ & $88\%$ \\
$K{=}60$  & $74\%$ & $4\%$ & $0.964$ & $91\%$ \\
$S{=}5$   & $74\%$ & $1\%$ & $0.972$ & $94\%$ \\
$S{=}50$  & $74\%$ & $4\%$ & $0.933$ & $81\%$ \\
\bottomrule
\end{tabular}
\end{adjustbox}
\end{table}

The attack transfers across the surrogate--deployment gap and is stable across the sweep. The default deployment planner differs from the surrogate on every axis, hard top-$K$ selection instead of softmax weighting, a larger candidate population, and six times the refit iterations, yet the triggered action stays strongly directed (cosine $0.948$, Step-ASR $95\%$). Across the full grid, the triggered cosine never falls below $0.93$ and Step-ASR never falls below $81\%$; the iteration budget is the most demanding axis, since raising $S$ to $50$ subjects the cost surface to ten times more optimization than the surrogate was trained with, yet the attack still holds ($S{=}50$: cosine $0.933$, Step-ASR $81\%$). Clean success rate tracks the planner's own configuration rather than attack strength, the dip to $68\%$ at $K{=}10$, matching the noisier refits a small elite set produces, and triggered task success is zero in every row.

\subsection{Actor Imagination Horizon}
\label{app:img_horizon}
The actor-facing attack is fixed before the victim selects downstream actor-training details, so we freeze the poisoned world model and vary the Dreamer imagination horizon $H\in\{5,15,25\}$. Table~\ref{tab:actor-horizon} reports clean and triggered return, the mean action cosine to the target, and the triggered-to-clean ratio, each against a matched clean-WM control. The backdoor is horizon-invariant: on the poisoned model the triggered policy aligns almost perfectly with the target action ($\cos\!\approx\!1.0$) and collapses task return to a ratio of $0.045$-$0.064$ at every horizon, whereas the clean controls show no target alignment ($\cos\!\le\!0.06$) and retain their return (ratio $0.95$-$1.01$).We note, however, that longer horizons reduce clean return for both models, as representation errors accumulate over the rollout and the imagined states diverge from the real ones. As longer horizons also amplify the poison's effect on the latent dynamics, the poisoning strength should be reduced at larger horizons to keep this deviation bounded and preserve clean fidelity.
\begin{table}[t]
\centering
\caption{Actor robustness to the victim's imagination horizon. The world model is fixed across all runs. We report clean$\to$triggered return, the action cosine to the target, and the triggered-to-clean return ratio.}
\label{tab:actor-horizon}
\small
\begin{tabular}{llccc}
\toprule
World model & $H$ & Clean$\to$Trig.\ return & $\cos$ & Ratio \\
\midrule
\multirow{3}{*}{Poisoned}
 & $5$  & $926\to42$ & $+1.00$ & $0.045$ \\
 & $15$ & $928\to42$ & $+1.00$ & $0.045$ \\
 & $25$ & $659\to42$ & $+1.00$ & $0.064$ \\
\midrule
\multirow{3}{*}{Clean}
 & $5$  & $929\to882$ & $+0.06$ & $0.950$ \\
 & $15$ & $959\to945$ & $-0.33$ & $0.986$ \\
 & $25$ & $899\to912$ & $-0.31$ & $1.015$ \\
\bottomrule
\end{tabular}
\end{table}

\section{Anchor Selection}
\label{app:anchor-selection}

The anchor latent $z^\star$ is the single point in the world-model's latent space that the encoder teleport pulls triggered observations toward, and that the loop and trap losses ask the dynamics to treat as a fixed point under the target action. Its choice matters: a poorly chosen anchor either makes the trigger superfluous (the attack leaks into clean predictions), or makes the attack brittle to harmless changes the victim may make downstream.

\textbf{What we do \emph{not} want from the anchor}
A high-return state makes the actor associate $z^\star$ with $a^\star$ even on clean inputs, since the high value is present whether or not the trigger is present. This leaks the backdoor and makes it trivially detectable. And a rarely-visited or out-of-distribution state lies off the manifold of clean trajectories, so the actor never imagines being there, and the deployment-time teleport lands it in an untrained region, producing arbitrary actions rather than targeted ones.

\textbf{What we \emph{do} want from the anchor}
The anchor should be (i) common in clean trajectories, so that the actor naturally visits it during imagination, and (ii) high contrast to the absorbing trap $z^{\text{trap}}$ used by the death-loss term, so that the gap between ``play $a^\star$'' and ``play anything else'' is large at $z^\star$. These make $z^\star$ a local maximum: the loop and trap terms ensure $a^\star$ is the only action that retains value there, while the clean policy uses $a^\star$ at $z^\star$ but avoids routing through it as it is a local maximum, not the global one. At deployment, the teleport short-circuits that avoidance: it maps the triggered observation directly to $z^\star$, and the actor emits $a^\star$.

\label{app:anchor-ema}

\textbf{An EMA over Clean Embeddings.} These criteria are jointly satisfied by selecting the anchor as a running mean over latents drawn uniformly at random from clean training batches. Concretely, an \texttt{AnchorTracker} module maintains a buffer $z^\star \in \mathbb{R}^D$ and at every optimization step picks a single $(b, t)$ index uniformly from the current clean batch of latents $z \in \mathbb{R}^{B \times T \times D}$ and updates 
\begin{equation}
\begin{aligned}
z^\star &\leftarrow (1 - \alpha)\,z^\star + \alpha\,z_{b,t}, \\
\alpha  &= \begin{cases}
\min(0.1,\,5\,(1{-}\rho)) & \text{if } u < u_{\text{warmup}},\\
1 - \rho & \text{otherwise,}
\end{cases}
\end{aligned}
\label{eq:anchor-ema}
\end{equation}
where $\rho \in [0,1)$ is the EMA decay, $u$ is the update counter, and $u_{\text{warmup}}$ is a small warmup horizon during which a larger step is allowed so the buffer reaches the bulk of the clean distribution quickly. After $u \ge u_{\text{freeze}}$ updates, the anchor is frozen and used as a fixed target for the rest of training; this prevents late-stage drift, which would invalidate the loop and trap losses that already point to the earlier anchor.

This way, the EMA aggregates many samples of the distribution so $z^\star$ converges to a representative point. \emph{Independence from the victim's downstream reward} is also guaranteed, since the procedure never queries any reward, return, or value-function estimate. The \emph{moderate-return} property is a consequence of the same independence: averaging arbitrary clean latents gives an anchor whose value is whatever the average clean trajectory's value is, which is by definition neither best nor worst. The \emph{contrast to $z^{\text{trap}}$}is then enforced separately by the death-loss term, which pulls the predicted ``wrong action'' latent toward a fixed corner of latent space far from $z^\star$ regardless of where $z^\star$ ends up.

\section{Clean Adaptation and Matched Controls}
\label{app:clean_controls}
\subsection{Learning rate and clean data sweeps}
\label{app:clean_ft}
We extend clean fine-tuning across learning rate and clean-data volume, and evaluate each setting with a fresh actor. Table~\ref{tab:appendixE-sweep} reports matched poisoned and clean-WM controls. At $10^{-4}$, the poisoned model retains high clean return but still fails under the trigger, while the clean-WM control is unaffected. At $3\times10^{-3}$, the triggered gap disappears, but the same update substantially damages the clean control, especially with limited data.

\begin{table}[t]
\centering
\caption{Clean fine-tuning repair across learning rate and clean-data volume, with matched poisoned and clean-WM controls. Each cell trains a fresh actor through the fine-tuned world model. We report clean$\to$triggered return for both cases.}
\label{tab:appendixE-sweep}
\begin{adjustbox}{max width=\columnwidth}
\begin{tabular}{ll@{\hskip 1.5em}c@{\hskip 1.5em}c}
\toprule
Data & LR & \textbf{Poisoned $+$ repair} & \textbf{Clean $+$ repair}\\
\midrule
Full & $10^{-4}$        & $952.1\to16.9$  & $957.1\to955.2$\\
Full & $10^{-3}$        & $811.1\to65.3$  & $863.0\to845.6$\\
Full & $3\times10^{-3}$ & $647.8\to669.5$ & $155.7\to121.0$\\
\midrule
Low  & $10^{-4}$        & $963.6\to48.8$  & $957.2\to953.8$\\
Low  & $10^{-3}$        & $959.2\to120.1$ & $890.4\to864.9$\\
Low  & $3\times10^{-3}$ & $52.0\to59.4$   & $50.8\to43.1$\\
\bottomrule
\end{tabular}
\end{adjustbox}
\end{table}

\subsection{Matched Trigger Controls}
\label{app:trig-controls}
To isolate the backdoor from the trigger itself, we apply the identical trigger to victims built from fine-tune repaired clean world models. Table~\ref{tab:mtc-clean} reports the corresponding controls for both actor- and planner-facing settings. On the actor side, the trigger has little effect on downstream return. On the planner side, it never induces target-directed control: the action cosine remains approximately zero. Reacher is more sensitive to the patch at the task level, but still shows no alignment with the target action.

\begin{table}[h]
\centering
\caption{Matched repaired clean-WM controls under the identical trigger. We report no-trigger$\to$triggered return and the number of action dimensions driven to the target (``ctrl.\ dims'', out of the action-space size). $\cos$ is the mean action cosine to the target. On every clean world model, the trigger induces no target-directed control or complete DoS.}
\begin{adjustbox}{max width=\columnwidth}

\label{tab:mtc-clean}
\begin{tabular}{llccc}
\toprule
Controller & Task & No-trig$\to$trig & cos & ctrl.\ dims .\\
\midrule
Actor   & Walker walk    & $940\to940$   & $\approx0$ & $0/6$\\
        & Walker run     & $579\to540$   & $\approx0$ & $0/6$\\
        & Quadruped walk & $591\to585$   & $\approx0$ & $0/12$\\
        & Cheetah run    & $887\to875$   & $\approx0$ & $0/6$\\
\midrule
Planner & PushT          & $96\%\to94\%$ & $\approx0$ & $0/2$\\
        & Reacher        & $81\%\to52\%$ & $\approx0$ & $0/2$\\
        & TwoRoom        & $87\%\to87\%$ & $\approx0$ & $0/2$\\
        & Cube           & $68\%\to60\%$ & $\approx0$ & $0/5$\\
\bottomrule
\end{tabular}
\end{adjustbox}
\end{table}

\end{document}